\documentclass[11pt]{article}

\usepackage[utf8]{inputenc}
\usepackage[T1]{fontenc}
\usepackage{lmodern}
\usepackage{microtype}
\usepackage[a4paper,margin=2.5cm]{geometry}
\usepackage{amsmath,amssymb,mathtools}
\usepackage{graphicx}
\usepackage{booktabs,tabularx,array}
\usepackage{caption}
\usepackage{float}
\usepackage{xcolor}
\usepackage{pdflscape}
\usepackage{xurl}
\usepackage[hidelinks]{hyperref}
\usepackage{ragged2e}
\usepackage{soul}

\newcolumntype{Y}{>{\RaggedRight\arraybackslash}X}
\newcolumntype{P}[1]{>{\RaggedRight\arraybackslash}p{#1}}
\DeclareCaptionLabelSeparator{bar}{\space\textbar\space}
\newcommand{\placeholderfigure}[1]{%
  \IfFileExists{#1.pdf}{%
    \includegraphics[width=\textwidth]{#1.pdf}%
  }{\IfFileExists{#1.png}{%
    \includegraphics[width=\textwidth]{#1.png}%
  }{\IfFileExists{#1.jpg}{%
    \includegraphics[width=\textwidth]{#1.jpg}%
  }{%
    \fbox{\parbox[c][0.25\textheight][c]{0.92\textwidth}{%
      \centering\textbf{#1}\\[0.5em]\emph{Figure placeholder}%
    }}%
  }}}%
}

\title{Data geometry preserves prediction but reshapes explanation}

\author{Nicola Amoroso\textsuperscript{1,2,\textdagger},
Mario Caruso\textsuperscript{4,\textdagger},
Marianna La Rocca\textsuperscript{4,3,*},
Loredana Bellantuono\textsuperscript{3,4},\\
Tommaso Maggipinto\textsuperscript{3,3},
Michele Morelli\textsuperscript{4},
Sabina Tangaro\textsuperscript{3,6},
Marco Tatullo\textsuperscript{5},\\
Roberto Bellotti\textsuperscript{4,3},
Ester Pantaleo\textsuperscript{4,3,\textdaggerdbl},
Alfonso Monaco\textsuperscript{4,3,\textdaggerdbl}
\\[1em]
\small
\textsuperscript{1}Dipartimento di Farmacia -- Scienze del Farmaco,\\
\small
Università degli Studi di Bari Aldo Moro, Bari, Italy
\\
\small
\textsuperscript{2}rAIse s.r.l.,Bari, Italy\\
\small
\textsuperscript{3}Istituto Nazionale di Fisica Nucleare,\\
\small
Sezione di Bari, Bari, Italy
\\
\small
\textsuperscript{4}Dipartimento Interuniversitario di Fisica ``M. Merlin'',\\
\small
Università degli Studi di Bari Aldo Moro, Bari, Italy
\\
\small
\textsuperscript{5}Dipartimento di Biomedicina Traslazionale e Neuroscienze,\\
\small
Università degli Studi di Bari Aldo Moro, Bari, Italy\\
\small
\textsuperscript{6}Dipartimento di Scienze del Suolo, della Pianta e degli Alimenti,\\
\small
Università degli Studi di Bari Aldo Moro, Bari, Italy
\\[0.8em]
\footnotesize
\textsuperscript{\textdagger}Equal first-author contribution.
\\
\footnotesize
\textsuperscript{\textdaggerdbl}Equal last-author contribution.
\\
\footnotesize
\textsuperscript{*}Correspondence to Marianna La Rocca:
\texttt{marianna.larocca@uniba.it}
}

\date{}

\begin{document}
\maketitle
\section{Abstract}\label{abstract}

Across scientific domains, empirical data often give rise to positive semidefinite matrices that encode similarities, couplings or interactions and induce natural geometries. We investigate whether changing the geometry used to compare the same empirical representations preserves subject-level performance and the local structures from which explanations are derived, a question we term the prediction–explanation invariance problem. We address this problem across morphometric MRI, fMRI and EEG data by comparing Frobenius and trace geometries applied to the same empirical matrices. Predictive performance was broadly preserved across geometries, but comparable performance did not imply agreement in subject-level decision scores or predicted labels. Predictive similarity also masked geometry-dependent differences in local neighbourhoods, perturbation sensitivities and explanatory rankings. Dimensionality reduction made the two subject-space geometries progressively more concordant, while classification performance and decision-level agreement declined. These results identify data geometry as a hidden degree of freedom in explainability: similar performance does not guarantee invariant decisions or explanations.

\textbf{Keywords:} explainable machine learning; data geometry; density
operators; positive semidefinite matrices; trace distance; Frobenius
geometry; representation invariance; prediction–explanation invariance.

\section{The prediction-explanation invariance
problem}\label{the-prediction-explanation-invariance-problem}

Machine-learning models are commonly evaluated by their ability to make accurate predictions. Yet in scientific, clinical and regulatory settings, explainability is increasingly essential: a slightly less accurate but interpretable model may be preferable to a more accurate model whose decisions cannot be inspected \cite{jimenez2020drug,amann2020explainability,amann2022explain,afroogh2024trust,anang2024explainable,papadakis2024explainable,nicolson2025human}. When two representations or algorithms achieve similar predictive performance, they are therefore often treated as capturing comparable information about the problem. This assumption is convenient but incomplete. Predictive similarity can be considered at two complementary levels: aggregate performance, which measures how much discriminative information a representation retains, and decision agreement, which measures whether alternative models assign similar scores and labels to the same observations.

Prediction and explanation nevertheless address different properties of a representation. Prediction asks whether it contains sufficient information to separate outcomes and whether alternative representations support concordant individual decisions. Explanation asks which components are most sensitive to local perturbations and which local structures support the observed organization of the data. Two representations may therefore achieve comparable performance while inducing different decision-level and explanatory structures \cite{dong2020exploring,d2022underspecification,lombardi2022reliability}. We define the prediction–explanation invariance problem as the question of whether changing the geometry used to compare the same empirical representations preserves predictivity, at both aggregate and subject-specific levels, together with explanatory structures derived from neighbourhoods and node perturbations.

A central but often implicit source of such variability is the geometry used to compare data \cite{dombrowski2019explanations,zaher2024manifold}. Geometry affects distances, similarities, margins, kernel structures and neighbourhood organization, and is known to influence learning from covariance and positive-definite matrix representations \cite{pennec2006riemannian,barachant2012riemannian,congedo2017riemannian}. Its role in explanation is less well understood. Geometry determines which observations are considered close, which perturbations are regarded as large or small and which local changes are amplified or suppressed. This is particularly relevant because many attribution and perturbation-based methods infer importance from structured modifications of the input or its local context \cite{ribeiro2016lime,lundberg2017shap,agarwal2021unification,vu2025emap,ghorbani2019fragile,yeh2019infidelity}. If locality depends on geometry, explanation cannot be assumed to be invariant to the metric used to analyse the data.

Positive semidefinite matrices provide a natural setting in which to examine this problem. They admit Gram representations and can be interpreted as kernel- or covariance-type objects with non-negative spectra and an implicit embedding in a Euclidean or Hilbert space \cite{scholkopf2002kernels,bhatia2007positive}. Such matrices arise from similarities, correlations, covariances and coupling patterns across scientific domains. In brain data, for example, they may encode similarities between spatial intensity profiles, haemodynamic coupling between brain regions or electrophysiological coupling between channels \cite{friston1994functional,van2010intrinsic,sporns2011human}. Their common algebraic structure makes it possible to change the geometry while keeping the empirical representation fixed.

We compare two complementary geometries on the same positive semidefinite matrices. Frobenius geometry is a classical quadratic, (L\textsubscript{2})-like matrix geometry that treats differences as distributed changes across entries. Trace geometry is an operator, (L\textsubscript{1})-like geometry that compares matrices through the directions along which they can be distinguished and corresponds to a standard distinguishability measure for density operators \cite{nielsen2010quantum}. Because both geometries are applied to identical matrices and evaluated through the same classifier architecture, differences in neighbourhood structure or perturbation-based importance reflect how the selected metric propagates the same node-level changes through the corresponding kernel and decision function, rather than differences in the underlying data.

We address the prediction–explanation invariance problem across three regimes: morphometric MRI similarities between spatial intensity profiles, fMRI haemodynamic coupling between anatomical regions and EEG electrophysiological coupling between channels \cite{rykhlevskaia2008combining,park2013structural,abdelnour2014network,rasero2017structural,larocca2017multiplex,amoroso2018multiplex,van2010exploring,smith2011network,ma2016resting,chen2008dynamic,cohen2008assessing}. For each geometry, predictive behaviour was evaluated through both held-out performance and agreement in subject-level decision scores and predicted labels. Explanatory structure was assessed separately through subject-space neighbourhoods and node-level perturbation rankings. A controlled node-subsampling analysis further tested how these forms of agreement changed as representation dimensionality and discriminative content were progressively reduced. This design distinguishes predictive similarity, decision agreement and explanatory invariance, and tests whether geometry acts merely as an alternative computational lens, or whether it changes the organization of the subject space, thus  constituting a hidden degree of freedom in explainable machine learning.

\section{Results}\label{results}

The prediction–explanation invariance problem was examined across three empirical domains that generate positive semidefinite subject-level matrices with distinct interpretations (Table~\ref{tab:table1}).

\begin{table}[H]
\centering
\small
\begin{tabularx}{\textwidth}{@{}l l c Y Y@{}}
\toprule
\textbf{Dataset} & \textbf{Subjects (groups)} & \textbf{Nodes} & \textbf{Node definition} & \textbf{Matrix entries} \\
\midrule
Morphometric MRI & 96 AD, 119 CTRL & 549 & Spatial image patches & Pearson similarity between patch intensity profiles \\
fMRI & 403 ASD, 468 CTRL & 116 & AAL anatomical regions & Pearson correlation between regional haemodynamic time series \\
EEG & 36 AD, 29 CTRL & 19 & Scalp EEG channels & Pearson correlation between electrophysiological time series \\
\bottomrule
\end{tabularx}
\caption{Empirical regimes and subject-level positive semidefinite representations. The three datasets define distinct empirical regimes that generate subject-level positive semidefinite matrices. Morphometric MRI matrices encode spatial similarity between image-patch intensity profiles, fMRI matrices encode haemodynamic coupling between anatomical regions, and EEG matrices encode electrophysiological coupling between scalp channels. In all datasets, the resulting subject-level matrices were symmetrized and trace-normalized to obtain a common positive semidefinite representation for Frobenius and trace geometry analyses.}
\label{tab:table1}
\end{table}

\begin{figure}[!htbp]
\centering
\includegraphics[width=0.95\textwidth]{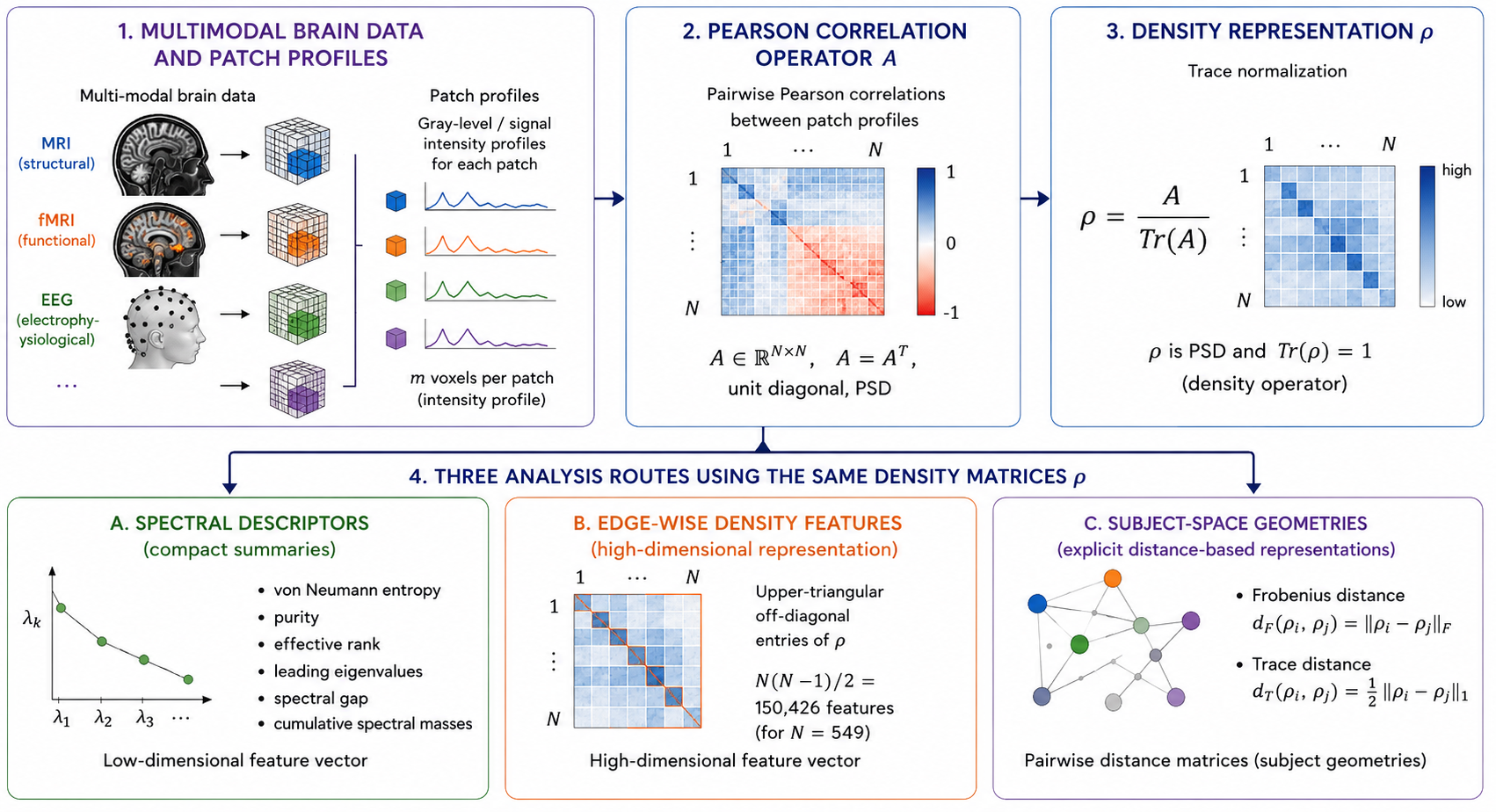}
\caption{Brain data from different modalities, including structural MRI, functional MRI, and EEG, are represented through modality-specific regional or patch-level profiles. Pairwise Pearson correlations between these profiles define a subject-specific correlation operator $A$, whose full correlation form has unit diagonal and is positive semidefinite. Trace normalization yields the density representation $\rho=A/Tr(A)$. The same density matrices are then analyzed through compact spectral descriptors, high-dimensional edge-wise density features, and explicit subject-space geometries based on Frobenius and trace distances. Predictive analyses test the disease-related informative power of each representation, whereas perturbation-based explainability compares the patch-level sensitivities induced by the two geometries.}
\label{fig:figure1}
\end{figure}

Morphometric MRI data were represented as patch-by-patch similarities
between spatial intensity profiles; fMRI data as region-by-region
haemodynamic coupling matrices; and EEG data as channel-by-channel
electrophysiological coupling matrices. A detailed description of
dataset sources, inclusion criteria and preprocessing choices is included in Supplementary Note 1. For each dataset, the same analysis pipeline compared compact spectral descriptors, edge-wise matrix features, Frobenius geometry and trace geometry. Informative content was assessed using nested cross-validation. Subject-space organization was then examined through global distance concordance and local neighbourhood overlap, whereas perturbation-based explanatory structure was assessed through node-level sensitivities and rankings. These analyses avoided introducing a separate post-hoc feature-attribution algorithm and isolated how alternative geometries propagated the same structured node perturbations through an otherwise matched classifier architecture. This design allowed predictivity and explanatory invariance to be evaluated on the same empirical matrices, rather than across different input data or preprocessing pipelines (Fig.~\ref{fig:figure1}).

\subsection{Full-matrix geometries retain predictive
information}\label{full-matrix-geometries-retain-predictive-information}

Predictive information was first evaluated by comparing four
representation families. Compact spectral descriptors provided a
low-dimensional summary of each matrix and served as a reduced global
baseline. Edge-wise matrix features were used as a full-matrix
benchmark, preserving all pairwise entries without imposing an explicit
subject-space geometry. Frobenius and trace representations instead used
the same matrices to define two alternative geometries between subjects.

\begin{figure}[H]
\centering
\includegraphics[]{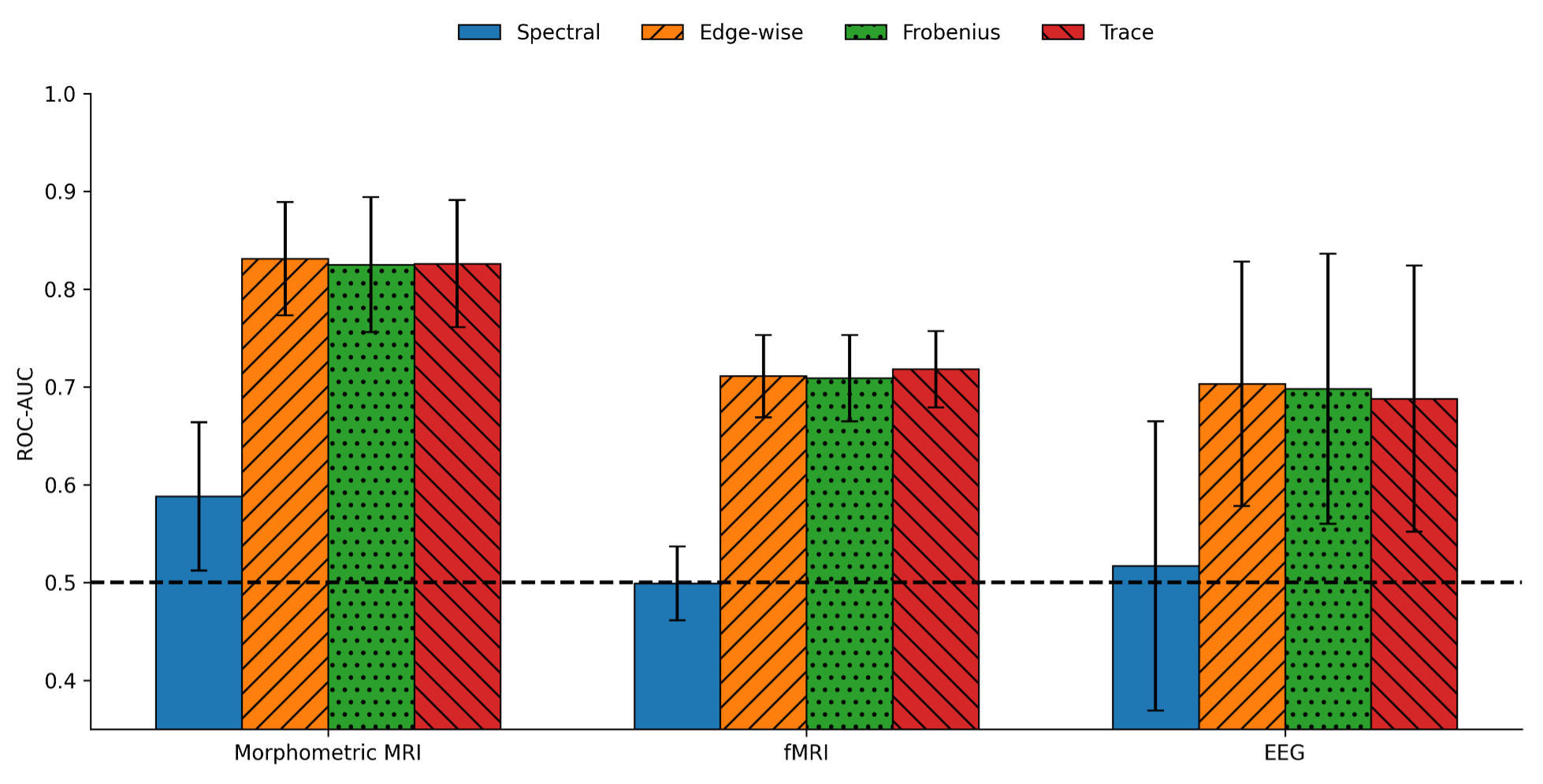}
\caption{Full-matrix geometries retain predictive information across empirical regimes. Predictive performance of compact spectral descriptors, edge-wise full-matrix features, Frobenius geometry and trace geometry across morphometric MRI, fMRI and EEG datasets. Bars show mean ROC-AUC across outer cross-validation evaluations, with error bars indicating standard deviation. The dashed line denotes chance-level performance. Compact spectral summaries were weakly informative across datasets, whereas edge-wise features and geometry-based representations retained predictive information. Frobenius and trace geometries showed closely matched predictive performance when applied to the same empirical matrices.}
\label{fig:figure2}
\end{figure}

Compact spectral descriptors provided limited informative content across all three datasets, indicating that low-dimensional eigenvalue summaries did not capture the discriminative structure contained in the full matrices. Full-matrix representations were consistently more
informative. In morphometric MRI, edge-wise density features, Frobenius
geometry and trace geometry all achieved substantially higher
discrimination than spectral summaries, with ROC-AUC values around $0.83$.
In fMRI, the same pattern was observed, with edge-wise, Frobenius and
trace representations yielding ROC-AUC values in the $0.71-0.72$ range. In EEG,
full-matrix representations also retained predictive signal, with
ROC-AUC values ranging from $0.69$ to $0.70$, although geometry-based classifiers
showed lower specificity than edge-wise classifiers (Fig.~\ref{fig:figure2}).

Frobenius and trace geometries therefore preserved broadly comparable predictive information across domains and scales. This was a necessary condition for testing explanatory invariance: if one geometry had failed to retain task-relevant information, differences in explanation could have been attributed to poorer prediction rather than to geometry. Instead, the two geometries provided similar predictive access to the same empirical matrices, motivating a direct comparison of the local structures and explanations induced by each metric. Comparable performance, however, does not establish that the corresponding classifiers assign the same decision scores or labels to individual subjects; this distinction was examined explicitly in the dimensionality analysis.

To assess whether these findings depended on the choice of learning algorithm, classifier-specific performance was examined across the outer folds of the nested cross-validation procedure. The reduced performance of spectral descriptors and the stronger performance of full-matrix representations were consistent across the classifiers considered; detailed model-specific results are reported in Supplementary Note S3.

\subsection{Comparable predictive performance does not imply neighbourhood
invariance}\label{predictive-similarity-does-not-imply-neighbourhood-invariance}

Comparable predictive performance did not imply that Frobenius and trace geometries organized the subject space in the same way.  Subject-space concordance was quantified by first computing, for every pair of subjects, the Frobenius and trace distances between their density matrices and then assessing the Spearman correlation between the two resulting sets of pairwise distances. Concordance varied markedly across datasets.

Frobenius and trace distances were nearly collinear in EEG, with a
Spearman correlation of 0.98, and remained strongly related in
morphometric MRI, with a Spearman correlation of 0.91. In fMRI, however,
subject-space concordance was substantially lower, with a Spearman
correlation of 0.62. Thus, the same empirical matrices could support
nearly equivalent or substantially different subject-space organizations
depending on the domain and dimensional regime.

This divergence was more pronounced when local neighbourhoods were
compared directly. Local neighbourhood concordance was assessed by identifying, for each subject and within each geometry, the $k$ nearest subjects according to the corresponding pairwise distances. The overlap between the Frobenius- and trace-based neighbourhoods was then quantified using the Jaccard index and averaged across subjects.

\begin{figure}[H]
\centering
\includegraphics[scale=0.4]{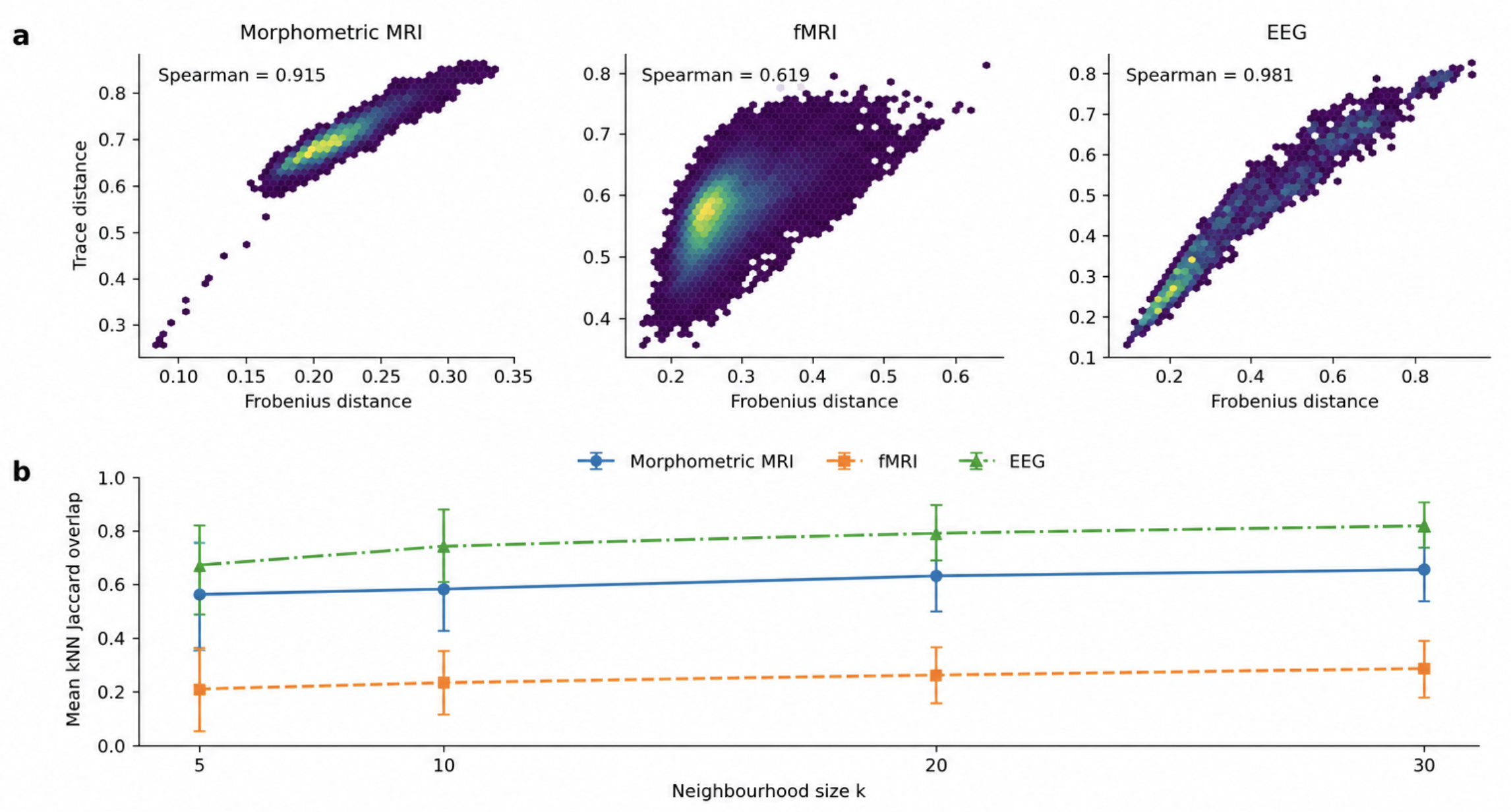}
\caption{Predictive similarity does not imply neighbourhood invariance. a) For each dataset, pairwise Frobenius and trace distances were computed between all subject-specific density matrices. Global subject-space concordance was quantified as the Spearman correlation between the two resulting sets of pairwise distances. b) Local neighbourhood concordance was assessed by identifying, for each subject and within each geometry, the k nearest neighbours according to the corresponding pairwise distances. The overlap between Frobenius- and trace-based neighbourhoods was quantified using the Jaccard index and averaged across subjects for $k=5$, $10$, $20$, and $30$. Error bars indicate variability across subjects. Morphometric MRI showed high global but only moderate local concordance, fMRI showed the lowest concordance at both levels, and EEG showed consistently high global and local agreement.}
\label{fig:figure3}
\end{figure}

Across $k=5$, $10$, $20$, and $30$, mean neighbourhood overlap increased moderately with $k$ in all datasets but remained consistently highest in EEG, intermediate in morphometric MRI, and lowest in fMRI. The mean Jaccard index increased from $0.68$ to $0.82$ in EEG, from $0.56$ to $0.67$ in morphometric MRI, and from $0.20$ to $0.28$ in fMRI. These results indicate that a high correlation between distance matrices does not necessarily imply preservation of local neighbourhood structure. They also show that predictive similarity and geometric invariance are distinct properties: two representations may retain comparable discriminative information while organizing the same subjects differently at both global and local levels (Fig.~\ref{fig:figure3}).

Because many explanation methods rely on local perturbations or local comparisons, neighbourhood-level divergence suggests a plausible mechanistic basis for geometry-dependent explanations. To assess whether the two distance spaces produced a simple unsupervised separation of the diagnostic classes, we compared mean distances between subject pairs belonging to the same diagnostic class with those between subjects belonging to different diagnostic classes. We further examined whether geometry-dependent neighbourhood changes merely reflected differences in the global separation of the diagnostic classes. Unlike the supervised analyses in the previous section, which assessed whether the full distance matrices retained information exploitable by a classifier, the present analyses evaluate the intrinsic organization of the subject space directly, without fitting a predictive model. Pairwise distance AUC values ranged from $0.503$ to $0.528$, and silhouette coefficients computed from the precomputed distance matrices were consistently close to zero, indicating weak unsupervised class separation across datasets and geometries. Hence, similar predictive performance cannot be interpreted as evidence of a common class-clustering structure. Rather, the geometries preserve diagnostically relevant information while inducing distinct local organizations of the subject space. This establishes subject-space geometry as a separate level of comparison between prediction and perturbation-based explanation.

\subsection{Geometry reshapes perturbation-based
explanations}\label{geometry-reshapes-perturbation-based-explanations}

\begin{figure}[!htbp]
\centering
\includegraphics[]{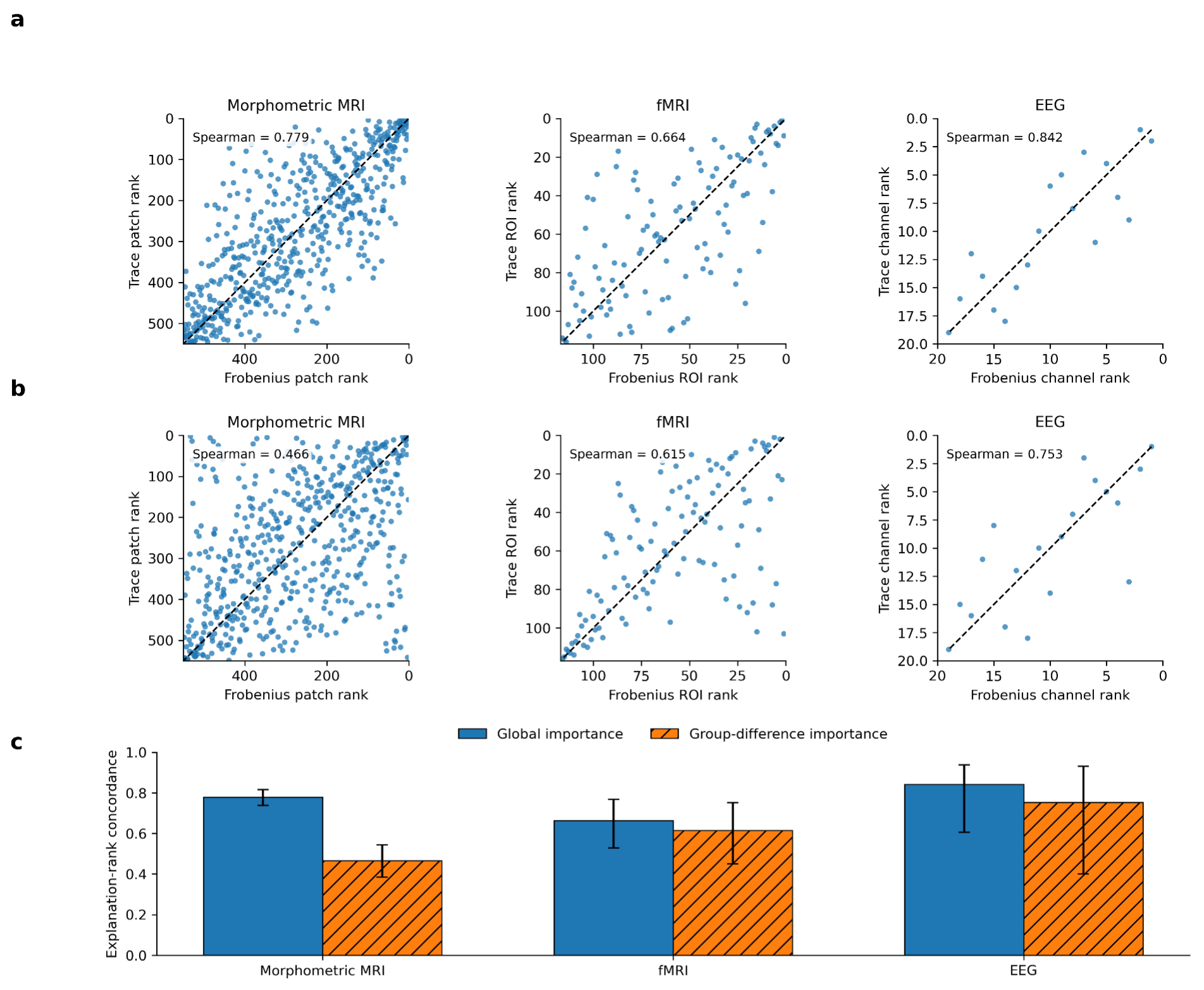}
\caption{Geometry reshapes perturbation-based explanations. a) Concordance between Frobenius- and trace-based global importance rankings across nodes for morphometric MRI, fMRI and EEG. Each point represents a patch, region or channel; lower ranks indicate higher perturbation-based importance. The dashed diagonal indicates identical rankings. b) Concordance between Frobenius- and trace-based group-difference importance rankings, quantifying whether the two geometries identify the same nodes as most characteristic of between-group explanatory differences. c) Summary of explanation-rank concordance. Bars show Spearman correlations between node-importance rankings obtained under Frobenius and trace geometries, separately for global importance and group-difference importance; error bars indicate 95\% bootstrap confidence intervals over nodes. The figure shows that geometry can leave global explanatory structure partly preserved while reshaping group-level explanatory rankings, especially in the morphometric MRI representation.}
\label{fig:figure4}
\end{figure}

Geometry-dependent differences in neighbourhood structure extended to perturbation-based explanations. For each dataset, the same structured perturbation was applied in turn to every patch, region or channel of each held-out subject. The perturbation modified that subject’s Frobenius- or trace-based distances from the training subjects and, consequently, the corresponding precomputed kernel values. Node importance was quantified as the absolute change in the decision score of the fitted geometry-based support-vector machine. These profiles should therefore be interpreted as geometry-conditioned decision explanations: perturbations were defined at the level of the original matrix representation, while their relevance was quantified through their effect on a classifier operating directly on the resulting geometry. Because the analyses were performed on the same empirical nodes and perturbations, they enabled a direct comparison of the explanatory structures induced by the two geometries (Fig.~\ref{fig:figure4}).

Frobenius- and trace-based rankings were not interchangeable. Global importance profiles showed partial concordance, together with substantial metric-specific structure. In morphometric MRI, the two geometries shared several high-ranking patches but differed markedly in the ordering and composition of the most influential regions. In fMRI, a common explanatory core was accompanied by pronounced geometry dependence in many local rankings. EEG rankings were more concordant, consistently with the near-collinearity of Frobenius and trace distances observed at the subject-space level.

The divergence was especially evident in the group-difference analyses. In morphometric MRI, Frobenius- and trace-based group-difference rankings showed only moderate concordance, indicating that the two geometries emphasized distinct disease-related perturbation patterns. In fMRI, concordance was stronger but remained incomplete, with both shared and metric-specific regional contributions. In EEG, group-difference rankings were more concordant, although still not identical. Thus, similar predictive performance does not imply explanatory equivalence: the same node perturbations can have different effects on the classifier decision function depending on the geometry through which they are propagated.

This pattern supports a geometry-conditioned interpretation of explanation divergence. Frobenius geometry propagates distributed entry-wise matrix changes, whereas trace geometry emphasizes operator-level directions along which matrices become distinguishable. Through the corresponding radial-basis kernels, these different perturbation responses alter the similarities between held-out subjects and the support vectors and may therefore produce different changes in the SVM decision function. Two geometries can consequently preserve comparable predictive performance while assigning different decision relevance to the same node perturbations.

Geometry therefore affects both the organization of the subject space and the perturbation-based explanatory structure derived from the same empirical representation. The main predictive, subject-space and explanation-rank results are summarized in
Table~\ref{tab:table2}.

\begin{table}[H]
\centering
\small
\begin{tabularx}{\textwidth}{@{}Y c c c@{}}
\toprule
\textbf{Measure} & \textbf{Morphometric MRI} & \textbf{fMRI} & \textbf{EEG} \\
\midrule
\textbf{Spectral ROC-AUC} & $0.588 \pm 0.076$ & $0.499 \pm 0.038$ & $0.517 \pm 0.148$ \\
\textbf{Edge-wise ROC-AUC} & $0.831 \pm 0.058$ & $0.711 \pm 0.042$ & $0.703 \pm 0.125$ \\
\textbf{Frobenius ROC-AUC} & $0.825 \pm 0.069$ & $0.709 \pm 0.044$ & $0.698 \pm 0.138$ \\
\textbf{Trace ROC-AUC} & $0.826 \pm 0.065$ & $0.718 \pm 0.039$ & $0.688 \pm 0.136$ \\
\textbf{Subject-space concordance} & $0.915\,[0.896, 0.934]$ & $0.619\,[0.590, 0.652]$ & $0.981\,[0.969, 0.987]$ \\
\textbf{kNN overlap, k = 10} & $0.586 \pm 0.163$ & $0.228 \pm 0.122$ & $0.751 \pm 0.137$ \\
\textbf{Global explanation-rank concordance} & $0.779\,[0.738, 0.816]$ & $0.664\,[0.528, 0.767]$ & $0.842\,[0.605, 0.938]$ \\
\textbf{Group-difference explanation-rank concordance} & $0.466\,[0.384, 0.543]$ & $0.615\,[0.449, 0.751]$ & $0.753\,[0.400, 0.931]$ \\
\bottomrule
\end{tabularx}
\caption{Predictive performance, subject-space concordance and explanation-rank concordance across datasets. Predictive performance is reported as mean $\pm$ standard deviation across outer cross-validation evaluations. Spectral ROC-AUC denotes the best compact spectral descriptor model; edge-wise ROC-AUC denotes the full-matrix benchmark; Frobenius and trace ROC-AUC values refer to geometry-based classifiers. Subject-space concordance is reported as the Spearman correlation between Frobenius and trace pairwise distance matrices, with 95\% bootstrap confidence intervals over subjects. kNN overlap is reported as mean $\pm$ standard deviation of Jaccard overlap across subjects at k = 10. Global and group-difference explanation-rank concordance values quantify the Spearman correlation between Frobenius- and trace-based perturbation-importance rankings, with 95\% bootstrap confidence intervals over nodes.}
\label{tab:table2}
\end{table}

Together, these results show that predictive similarity across
geometries can coexist with differences in subject-space organization
and perturbation-based explanatory rankings.

\section{Dimensionality modulates geometry, prediction and explanation}
To assess whether the dimensionality of the representation contributed to the observed differences between Frobenius- and trace-based geometries, we performed a controlled node-subsampling analysis in the morphometric MRI and fMRI datasets. EEG was not included because its representation already comprised a limited number of channels, and further dimensionality reduction would have produced very small systems with limited interpretability. 
At each retained-node fraction, from $90\%$ to $10\%$ of the original representation, random node subsets were sampled 30 times and applied identically to all subjects. Frobenius- and trace-based distances, neighbourhoods, unsupervised class-separation measures, classification performance, decision agreement and geometry-conditioned decision-importance rankings were then recomputed for each reduced representation.

\begin{figure}[!htbp]
\centering
\includegraphics[scale=0.45]{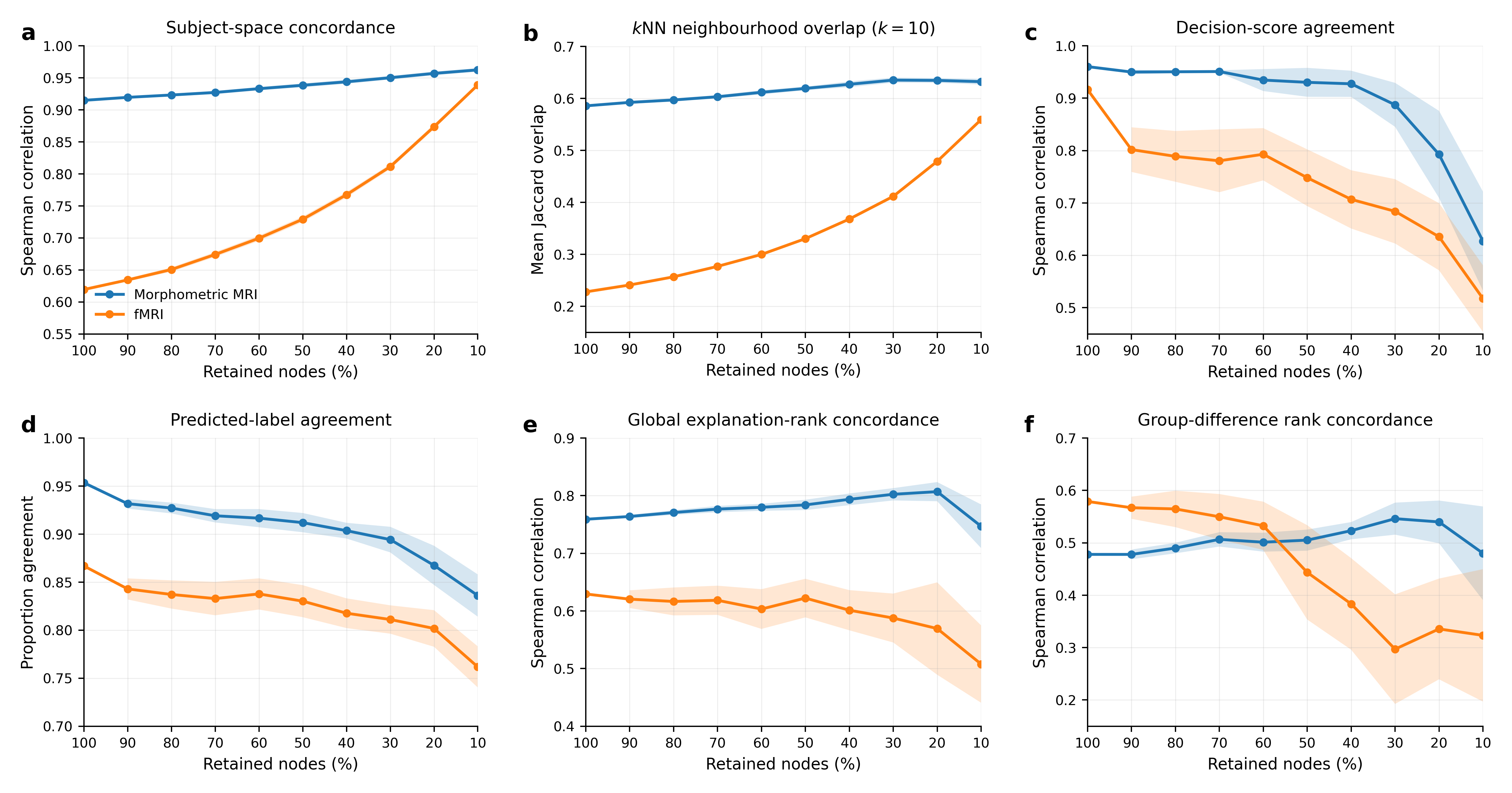}
\caption{Dimensionality differentially modulates geometric, decision-level and explanatory concordance. Node-dimensionality sensitivity analysis in fMRI and morphometric MRI. Random node subsets retaining $90\%$ to $10\%$ of the original representation were evaluated over 30 repetitions and applied identically to all subjects. a) Spearman correlation between vectorized Frobenius- and trace-distance matrices. b) Mean Jaccard overlap between the k=10 nearest-neighbour sets induced by the two geometries. c) Spearman correlation between out-of-fold decision scores. d) Proportion of subjects assigned the same predicted label. e) Spearman concordance between global perturbation-based importance rankings. f) Spearman concordance between group-difference perturbation rankings. Lines represent mean values and shaded areas indicate $95\%$ confidence intervals across random subsets. Full-dimensionality values were computed once. Lower dimensionality increased global and local geometric concordance, while decision-level agreement declined in both datasets and explanatory concordance followed dataset-dependent trajectories.}
\label{fig:figure5}
\end{figure}

Decreasing dimensionality consistently increased the concordance between the two subject-space geometries. In fMRI, the Spearman correlation between vectorized Frobenius- and trace-distance matrices increased progressively from $0.619$ at full dimensionality to $0.939$ when only $10\%$ of the nodes were retained. Morphometric MRI showed the same monotonic trend, although with a smaller absolute increase because the two geometries were already highly concordant at full dimensionality, rising from $0.915$ to $0.962$ (Fig.~\ref{fig:figure5}a). Local neighbourhood agreement followed the same general pattern. For $k=10$, the mean Jaccard overlap increased from $0.228$ to $0.559$ in fMRI. In morphometric MRI, it increased from $0.586$ to approximately $0.635$ at $20$–$30\%$ of the original nodes, before showing a slight decline to $0.632$ under the most aggressive reduction (Fig.~\ref{fig:figure5}b). The narrow confidence intervals indicated that these geometric and neighbourhood trends were highly stable across random node subsets. Pairwise distance AUC and silhouette coefficients remained essentially unchanged across dimensionality levels in both datasets, showing that the increasing agreement between Frobenius and trace was not driven by the emergence of more clearly separated diagnostic clusters.

Node removal progressively reduced the discriminative information retained by the representations. In morphometric MRI, Frobenius and trace ROC-AUC values decreased from approximately $0.82$ at full dimensionality to approximately $0.67$ when $10\%$ of the nodes were retained. In fMRI, ROC-AUC decreased from approximately $0.70$ to $0.56–-0.58$ over the same range. Decision agreement also declined despite the increasing concordance of the corresponding distance spaces. Spearman correlation between Frobenius- and trace-based out-of-fold decision scores decreased from $0.960$ to $0.627$ in morphometric MRI and from $0.917$ to $0.517$ in fMRI (Fig.~\ref{fig:figure5}c). Agreement between predicted labels decreased from $95.3\%$ to $83.6\%$ in morphometric MRI and from $86.7\%$ to $76.2\%$ in fMRI (Fig.~\ref{fig:figure5}d). Thus, lower-dimensional representations induced more similar global and local geometries, but progressively weaker and less concordant discriminative models.

The effect of dimensionality on perturbation-based explanations was instead modality-dependent. In morphometric MRI, moderate dimensionality reduction progressively increased global explanation-rank concordance (Fig.~\ref{fig:figure5}e): Spearman correlation rose from $0.759$ at full dimensionality to $0.807$ at $20\%$ of the nodes. Group-difference rankings showed a similar but weaker improvement, increasing from $0.478$ to approximately $0.54$–$0.55$ at $20$–$30\%$ of the original dimensionality (Fig.~\ref{fig:figure5}f). This convergence was not maintained under extreme reduction. At $10\%$ of the nodes, global concordance decreased to $0.747$, while group-difference concordance returned to approximately its full-dimensionality value and became substantially more variable across subsets.

A different pattern emerged in fMRI. Despite the strong and monotonic convergence of the underlying subject-space geometries and neighbourhoods, global explanatory concordance remained approximately stable through moderate reductions and then declined, with Spearman correlation decreasing from $0.629$ at full dimensionality to $0.508$ at $10\%$ of the nodes (Fig.~\ref{fig:figure5}e). The divergence was more pronounced for group-difference rankings, whose concordance decreased from $0.579$ to $0.323$ (Fig.~\ref{fig:figure5}f). Thus, lower dimensionality consistently reduced global and local geometric differences between Frobenius and trace, but did not produce a corresponding convergence of either subject-level decisions or perturbation-based explanations. Decision agreement declined in both datasets, whereas explanatory concordance increased under moderate reduction in morphometric MRI but remained stable or decreased in fMRI.

\section{Discussion}\label{discussion}

This study shows that comparable predictive performance does not guarantee either decision-level or explanatory invariance. Across three empirical regimes, Frobenius and trace geometries preserved broadly comparable discriminative information when applied to the same positive semidefinite matrices. However, they did not always induce the same neighbourhoods, decision scores, perturbation sensitivities or explanatory rankings. Predictive performance alone is therefore insufficient to establish the equivalence of two geometric representations. Geometry can affect the organization of the subject space, the decisions produced by the resulting classifiers and the perturbation-based relevance assigned to individual nodes. This distinction is consistent with the broader problem of underspecification, whereby systems with similar performance may nevertheless rely on different internal or local structures \cite{d2022underspecification,dong2020exploring,lombardi2022reliability}.

This result is especially relevant for explainable machine learning because many explanatory methods are local. Attribution, perturbation and neighbourhood-based analyses depend on how changes are measured and on which observations or directions are considered close to a given instance \cite{ribeiro2016lime,lundberg2017shap,agarwal2021unification}. Unlike conventional post-hoc feature attribution, the perturbations were not defined on the training-subject similarity coordinates used internally by the precomputed-kernel SVM. They were defined on interpretable nodes of the original positive semidefinite matrix and then propagated through the distance geometry, kernel and decision function. Frobenius geometry treats matrix differences as distributed entry-wise changes, whereas trace geometry emphasizes operator-level directions of distinguishability \cite{bhatia2007positive,nielsen2010quantum}. Both are natural for positive semidefinite data, but they weight the same perturbations differently. Because the kernel SVM depends explicitly on these distances, such differences directly modify the similarities to the support vectors and therefore the response of the decision function. The same empirical matrices can therefore support comparable performance while inducing different neighbourhoods, decision functions and perturbation-based explanatory structures.

The node-subsampling analysis clarified how dimensionality contributes to this distinction. In both morphometric MRI and fMRI, reducing the number of retained nodes progressively increased the concordance between Frobenius- and trace-distance matrices and the overlap between their local neighbourhoods. Lower-dimensional representations therefore produced increasingly similar descriptions of the subject space. At the same time, classification performance and agreement between out-of-fold decision scores and predicted labels declined. This apparent dissociation is not contradictory. Distance-matrix concordance summarizes the dominant organization of all subject pairs, whereas classification depends on the smaller component of that organization that carries information about group separation. Node removal can suppress metric-specific variability and make the overall geometries more similar while simultaneously eroding discriminative information.

Perturbation-based explanations followed a distinct pattern. In morphometric MRI, global and group-difference rankings showed nearly parallel trajectories: moderate dimensionality reduction increased their concordance between Frobenius and trace, whereas the strongest reduction reversed this tendency. This pattern is consistent with a regularization-like effect. When the discriminative signal is relatively robust, moderate node removal may eliminate weak or redundant components and leave a more compact informative structure that the two geometries rank more similarly. Once reduction becomes too severe, task-relevant components are also removed, performance deteriorates and explanatory rankings become less stable.

In fMRI, both global and group-difference explanation concordance decreased as dimensionality was reduced. Here the baseline discriminative signal was weaker, and node removal was associated with a progressive loss of information rather than an initial simplification benefit. Dimensionality reduction may therefore disrupt distributed and context-dependent functional relationships, making the relevance assigned to retained regions increasingly sensitive to the geometry used to evaluate the representation. The contrast between morphometric MRI and fMRI suggests that dimensionality reduction does not have a uniform effect on explanatory agreement. Its consequences depend on whether node removal primarily reduces redundancy or erodes task-relevant signal.

The EEG dataset could not provide an equivalent within-dataset test because its representation contained only 19 channels. Further proportional reduction would have generated very small systems and would not have offered a fair comparison with the MRI and fMRI trajectories. Its high subject-space and explanatory concordance should therefore remain a descriptive cross-modal observation rather than evidence that low dimensionality alone guarantees geometry-independent explanations. More generally, modality, sample size, noise structure, effective rank and the distribution of discriminative information remain partially confounded and should be manipulated independently in future controlled studies.

Taken together, these analyses refine the relation among geometry, prediction and explanation. Decreasing dimensionality can make two geometries more similar in their global and local descriptions of the subject space while making their classifiers less concordant at the level of individual decisions. Explanatory concordance may then either increase or decrease depending on the strength, redundancy and distribution of the retained signal. Geometry-dependent explanation should therefore be understood as the outcome of an interaction among the metric, the information content of the representation and its dimensionality, rather than as a simple consequence of predictive performance or matrix size alone.

These findings suggest that geometry choice should become an explicit component of model reporting. Explanations obtained under one metric may not be interchangeable with those obtained under another, even when performance is similar. Future work should compare families of geometries, distinguish random from anatomically or functionally structured subsampling, and manipulate independently discriminative signal, redundancy, effective rank and noise. Geometry-sensitivity diagnostics could then help determine whether an explanation reflects robust task-relevant structure or an implicit modelling choice \cite{adebayo2018sanity,dombrowski2019explanations,wang2020smoothed}. The trace-geometry perspective also provides a connection with quantum-information notions of distinguishability and motivates the exploration of operator divergences and measurement-inspired perturbations as tools for geometry-aware explanation \cite{dedomenico2016spectral,amoroso2021quantum}. In this view, geometry is not a technical detail of representation learning, but a modelling choice that shapes data organization, classifier behaviour and perturbation-based explanation. 

\section{Methods}\label{methods}

\subsection{Data sources and subject-level
matrices}\label{data-sources-and-subject-level-matrices}

Three empirical datasets were used to test whether geometry-dependent
explanations arise across distinct positive semidefinite
representations. The morphometric MRI dataset was derived from the
Alzheimer's Disease Neuroimaging Initiative (ADNI) \cite{jack2008adni} and included 215
subjects, comprising 96 patients with Alzheimer's disease and 119
cognitively normal controls. Each structural MRI volume was partitioned
into 549 spatial patches of size $10 \times 15 \times 20$ voxels. For each subject,
patch-wise grey-level intensity profiles were z-normalized and pairwise
Pearson correlations between patches were computed, yielding a 549 x 549
morphometric similarity matrix, consistently with previous subject-specific morphometric connectivity frameworks \cite{rasero2017structural,larocca2017multiplex,amoroso2018multiplex}.

The fMRI dataset was derived from the Autism Brain Imaging Data Exchange (ABIDE)
\cite{dimartino2014abide}, using derivatives distributed through the
Preprocessed Connectomes Project
\cite{craddock2013neurobureau}. It included 871 subjects, comprising 403 individuals
with autism spectrum disorder and 468 controls. Regional time series were extracted using the Automated Anatomical Labeling (AAL) atlas \cite{tzourio2002aal}, yielding 116 anatomical regions. For each subject, pairwise Pearson correlations between regional
haemodynamic time series were computed to obtain a 116 x 116 functional
coupling matrix.

The EEG dataset was derived from OpenNeuro ds004504 \cite{miltiadous2023eeg} and included 65
subjects after retaining Alzheimer's disease and cognitively normal
groups, comprising 36 patients and 29 controls. Scalp EEG recordings
were average-referenced, band-pass filtered between 1 and 45 Hz and
resampled to 250 Hz. For each subject, pairwise Pearson correlations
between 19 EEG channels were computed, yielding a 19 x 19
electrophysiological coupling matrix. These zero-lag channel correlations were treated as statistical coupling measures rather than direct estimates of neuronal interaction, given the known sensitivity of scalp-EEG connectivity to referencing, volume conduction, field spread and common sources \cite{bastos2016connectivity,vandiessen2015eeg}.

In all datasets, matrices were symmetrized before downstream analysis
and the diagonal was retained. Additional dataset-specific preprocessing
details, inclusion criteria and file-generation scripts are discussed in
Supplementary Note 1.

\subsection{Positive semidefinite
normalization}\label{positive-semidefinite-normalization}

For each subject, the empirical relation matrix was symmetrized and
checked for positive semidefiniteness by eigendecomposition \cite{bhatia2007positive}. Small
negative eigenvalues attributable to numerical precision were tolerated when negligible relative to the matrix scale. All matrices satisfied
positive semidefiniteness within numerical tolerance ($\varepsilon = 10^{-8}$).

Let $A_s$ denote the subject-level matrix for subject s. Because all
matrices had unit diagonal, their trace was fixed by the number of
nodes. A trace-normalized representation was obtained by dividing each
matrix by its trace. This yielded positive semidefinite matrices with unit trace, satisfying the algebraic definition of density operators \cite{nielsen2010quantum} (in an algebraic and information-theoretic sense), allowing the same empirical objects to be compared using
Frobenius and trace geometries. No thresholding or sparsification was
applied before geometric or predictive analyses.

A formal derivation of the density-operator representation, Frobenius
and trace distances, and perturbation-based sensitivity measures is
presented in Supplementary Note 2.

\subsection{Representation families}\label{representation-families}

Four representation families were compared for each dataset. Spectral
descriptors provided a compact global summary of each trace-normalized
matrix and were computed from its eigenvalue spectrum
\cite{dedomenico2016spectral}. Edge-wise matrix features were used as a
full-matrix benchmark by vectorizing the strictly upper-triangular,
off-diagonal entries of each trace-normalized matrix. This representation
preserved all pairwise relations but did not impose an explicit
subject-space geometry. Full definitions and dimensionalities of the
spectral and edge-wise representations are provided in Supplementary
Note S2.

Frobenius geometry was defined by computing pairwise distances between
trace-normalized matrices using the Frobenius norm. Trace geometry was
defined by computing pairwise trace distances between the same matrices.
These two distance matrices were then used as geometry-based
representations for modelling and neighbourhood analysis.
Thus, spectral descriptors, edge-wise features and the two geometries
were all derived from the same subject-level matrices, ensuring that
differences in prediction or explanation reflected representation and
metric choice rather than different input data.

\subsection{Predictive modelling}\label{predictive-modelling}

Predictive performance was evaluated separately for each dataset and
representation family. Classification was performed using repeated
nested cross-validation with 20 repeats of stratified five-fold outer
cross-validation \cite{varma2006bias,varoquaux2017decoders}. Model selection was performed within each training
split using stratified five-fold inner cross-validation. Hyperparameters
were selected by maximizing ROC-AUC in the inner loop and performance
was estimated on held-out outer folds.

Compact spectral descriptors were evaluated using L2-regularized
Logistic Regression, linear Support-Vector machines, radial-basis-function
Support-Vector machines and Random Forests. Because the edge-wise
representation was high-dimensional relative to sample size, edge-wise
density features were evaluated using the two regularized linear
classifiers, namely L2-regularized Logistic Regression and linear
Support-Vector machines. Geometry-based representations were evaluated
using Support-Vector machines with precomputed radial-basis-function
kernels derived from the Frobenius and trace distance matrices \cite{schoenberg1938metric}. Support-vector machines with precomputed kernels were selected because they provide the most direct link between the geometry induced by the distance metric and the classifier decision function. Unlike feature-based classifiers, they operate directly on pairwise similarities derived from the Frobenius or trace distances, allowing the effect of geometry itself to be isolated while avoiding additional representation-dependent transformations. The same outer folds were used across
representation families within each dataset to facilitate comparison.

Performance was summarized using ROC-AUC as the primary metric, with
balanced accuracy, accuracy, sensitivity and specificity reported as
secondary metrics. Summary values were computed across the 100 outer
held-out evaluations generated by the repeated nested cross-validation
procedure. Performance and agreement between subject-level model outputs were treated as complementary properties and were analysed separately in the node-dimensionality experiment described below.

\subsection{Subject-space concordance and neighbourhood
overlap}\label{subject-space-concordance-and-neighbourhood-overlap}

Spearman correlation was used to quantify subject-space concordance
because the analysis asked whether Frobenius and trace geometries
induced a similar rank ordering of subject pairs, rather than whether
their absolute distance values were linearly related. This rank-based
measure is directly relevant to local neighbourhoods, which depend on
the relative ordering of distances.

Local neighbourhood overlap was assessed using nearest-neighbour sets.
For each subject and each geometry, the $k$ nearest neighbours were
identified from the corresponding distance matrix. The overlap between
Frobenius and trace neighbourhoods was quantified using the Jaccard
index, and values were averaged across subjects. Neighbourhood overlap
was evaluated for multiple values of $k$, including $k$ = 5, 10, 20 and 30,
where applicable.

To assess whether distance spaces produced trivial diagnostic
clustering, unsupervised group separation was summarized by comparing
within-group and between-group distances, computing pairwise diagnostic
AUC from distances and estimating silhouette coefficients using the
precomputed distance matrices \cite{rousseeuw1987silhouettes}.

\subsection{Perturbation-based
explanations}\label{perturbation-based-explanations}

Unlike conventional post-hoc feature-attribution methods, perturbations were defined on anatomically or functionally interpretable nodes of the original positive semidefinite matrices, rather than on the training-subject similarity coordinates used internally by the precomputed-kernel classifier. Their relevance was quantified through the response of the fitted geometry-based model. The resulting analysis should therefore be interpreted as a geometry-conditioned perturbation explanation of the classifier decision, rather than as a classifier-independent measure of matrix sensitivity.

For each outer cross-validation split, a support-vector machine was fitted using a precomputed radial-basis-function kernel derived from either Frobenius or trace distances among the training subjects. Each node, corresponding to a morphometric patch, fMRI region or EEG channel, was then suppressed in turn for every held-out subject. This operation modified the distances between the held-out subject and the training subjects according to the selected geometry, while the fitted classifier and its hyperparameters were kept fixed. The perturbed distances were converted into a new kernel row and supplied to the same classifier.

Let $d_{\mathcal{G}}(s,t)$ and $d_{\mathcal{G}}^{(-i)}(s,t)$ denote, respectively, the original and node-$i$-suppressed distances between held-out subject $s$ and training subject $t$, for geometry $\mathcal{G} \in \{F,T\}$ for Frobenius and trace. The corresponding kernels were

$$
K_\mathcal{G}(s,t) = \exp{[-\gamma_\mathcal{G} d_\mathcal{G}(s,t)^2]}
$$

and

$$
K_\mathcal{G}^{(-i)}(s,t) = \exp{[-\gamma_\mathcal{G} d_\mathcal{G}^{(-i)}(s,t)^2]}
$$

Subject-specific node importance was defined as the absolute change in the SVM decision score,

$$
I_{i,\mathcal{G}}(s) = |f_\mathcal{G}(s) - f_\mathcal{G}^{(-i)}(s)|
$$.

Global importance was obtained by averaging $I_{i,\mathcal{G}}(s)$ across held-out subjects and cross-validation evaluations. Group-specific profiles were obtained by averaging the same quantity separately within the positive and negative classes. Group-difference importance was defined as the signed difference between the two group means. Frobenius- and trace-based importance profiles were compared using Pearson and Spearman correlations, top-$k$ overlap and consensus-ranking analyses.

For Frobenius geometry, node suppression was computed exactly by removing the corresponding row-and-column contribution from the squared Frobenius distance. For trace geometry, the change in distance was approximated using first-order matrix perturbation theory before being propagated through the radial-basis kernel and fixed SVM decision function. Exact and first-order trace-based rankings were compared in morphometric MRI and showed strong concordance. Full derivations are provided in Supplementary Note 2.

\subsection{Node-dimensionality analysis}
The effect of representation dimensionality was evaluated within the morphometric MRI and fMRI datasets. EEG was excluded because its original representation contained only $19$ channels and further proportional reduction would have generated very small systems with limited interpretability.

For each dataset, random node subsets retaining $90\%$ to $10\%$ of the original nodes were sampled at $10\%$ intervals. Thirty subsets were evaluated at each reduced dimensionality, whereas the full representation was evaluated once. Within each repetition, the same retained nodes were applied to every subject and to both geometries. Reduced matrices were subsequently normalized to unit trace.

For each node subset, Frobenius and trace distance matrices, subject-space concordance, nearest-neighbour overlap, pairwise diagnostic AUC, silhouette coefficients and perturbation-based importance rankings were recomputed. This within-dataset design allowed dimensionality to be varied while preserving subject composition and applying identical reductions to the two geometries.

To make repeated perturbation analyses computationally feasible across node subsets, the dimensionality experiment used a reduced 3-fold cross-validation configuration. The same configuration was applied at every dimensionality level, including the full-dimensionality reference; consequently, this reference serves as the internal baseline for the subsampling trajectories and may differ slightly from the primary full-sample estimates reported in Fig. \ref{fig:figure4} and Table \ref{tab:table2}.

\subsection{Decision-level agreement}
To distinguish performance from agreement between individual model outputs, Frobenius- and trace-based classifiers were compared at the subject level for every node subset. The two geometries used identical retained-node subsets and identical outer cross-validation partitions. Model selection was performed separately within the training data for each geometry.

Out-of-fold decision scores were collected for every subject. Agreement between Frobenius- and trace-based scores was quantified using Spearman and Pearson correlations. Agreement between predicted labels was summarized using the proportion of identical predictions and Cohen’s $\kappa$. The overlap between the sets of Support-vector subjects was quantified using the Jaccard index. ROC-AUC, accuracy and balanced accuracy were recomputed at each dimensionality level to assess the discriminative information retained by the reduced representations.

\subsection{Statistical reporting and
reproducibility}\label{statistical-reporting-and-reproducibility}

Cross-validation performance metrics were summarized across outer
held-out evaluations using means and standard deviations. Subject-space
concordance and neighbourhood-overlap measures were reported as
descriptive statistics computed from the full dataset within each
empirical regime. Perturbation-based explanation-rank concordance was
summarized using Pearson and Spearman correlations between Frobenius-
and trace-based importance profiles, together with top-k overlap of
explanatory rankings.

Bootstrap confidence intervals for subject-space concordance were
estimated over subjects, whereas confidence intervals for
explanation-rank concordance were estimated over nodes \cite{efron1993bootstrap}. Within the node-dimensionality analysis, results at each reduced dimensionality were summarized across the $30$ random node subsets using means and $95\%$ confidence intervals. The full-dimensionality values were computed once and therefore reported without across-subset uncertainty. All analyses were implemented using a common pipeline across datasets.

\section{Data availability}\label{data-availability}

The morphometric MRI data were obtained from the Alzheimer's Disease
Neuroimaging Initiative (ADNI). Resting-state fMRI data were obtained
from the ABIDE preprocessed initiative, and EEG data were obtained from
OpenNeuro ds004504. Access to ADNI data is governed by ADNI data-use
policies. ABIDE and OpenNeuro data are publicly available from their
respective repositories. Derived matrices, standardized metadata and
analysis outputs used to reproduce the main results will be made
available where permitted by the source data licences and data-use
agreements.

\section{Code availability}

Analysis code used to generate subject-level matrices, perform trace
normalization, compute Frobenius and trace geometries, run models and estimate perturbation-based explanation rankings is publicly available in the \textit{Zenodo} repository \cite{cm_zenodo}.

\section{Acknowledgements}\label{acknowledgements}
This work was supported by the Italian funding within the Budget MIUR, Dipartimenti di Eccellenza (Law 232, 11 December 2016), Quantum Sensing and Modelling for One-Health (QuaSiModO), CUP H97G23000100001; by the AI-OMICA Project (PiattaformA Integrata multi-OMICA di Intelligenza Artificiale per la gestione, la profilazione, l'analisi e la visualizzazione avanzata di dati in oncologia), CUP B19J25000290005; and by MUR (D.D. No. 307, PN RIC 2021--2027, Project ECHO TWIN), co-funded by the European Union (ERDF), Action 1.1.2 (CUP B99H26000290007), Action 1.1.3b (CUP B92F26000440005), and Action 1.4.3 (CUP B89J26003490005).

\section{Author contributions}\label{author-contributions}
N.A. M.L.R. and M.C.: Conceptualization, Methodology, Formal analysis, Writing original draft. M.M. Software. All other authors: Writing, review and editing. All authors read and approved the final version of the manuscript.

\section{Competing interests}\label{competing-interests}

The authors declare no competing interests.

\newpage

\bibliographystyle{sn-nature}
\bibliography{apssamp.bib}

\end{document}

\section*{Supplementary Information}
\addcontentsline{toc}{section}{Supplementary Information}

\begin{center}
\textit{Data geometry preserves prediction but reshapes explanation}
\end{center}

\setcounter{section}{0}
\renewcommand{\thesection}{S\arabic{section}}
\renewcommand{\thetable}{S\arabic{table}}
\renewcommand{\theequation}{S\arabic{equation}}

\section{Dataset sources and preprocessing details}
\label{supp:datasets}

This note provides dataset-level information supporting the empirical positive-semidefinite representations used in the main analyses. The three datasets were selected to span distinct regimes of empirical relation matrices: morphometric MRI patch similarities, resting-state fMRI haemodynamic coupling, and EEG electrophysiological coupling. For each dataset, subject-level relation matrices were generated before applying the common trace-normalization, geometry, prediction, and perturbation-analysis pipeline described in the Methods.

\noindent\textbf{Official dataset resources:}
\href{https://adni.loni.usc.edu/data-samples/adni-data/}{ADNI data portal};
\href{https://preprocessed-connectomes-project.org/abide/}{ABIDE Preprocessed};
\href{https://openneuro.org/datasets/ds004504}{OpenNeuro ds004504}.

\subsection{Morphometric MRI dataset: ADNI}
\label{supp:adni}

The morphometric MRI dataset was derived from the Alzheimer's Disease Neuroimaging Initiative (ADNI) \cite{jack2008adni}. The analysis retained subjects labelled as Alzheimer's disease and cognitively normal controls, yielding 215 subjects in total: 96 patients with Alzheimer's disease and 119 controls. Each structural MRI volume was partitioned into 549 spatial patches of size $10 \times 15 \times 20$ voxels. For each subject, the grey-level intensity profile of each patch was extracted and $z$-normalized. Pairwise Pearson correlations between patch intensity profiles were then computed, yielding a $549 \times 549$ patch-similarity matrix for each subject \cite{rasero2017structural,larocca2017multiplex,amoroso2018multiplex}.

Because the resulting matrix can be written as a Gram matrix of normalized patch profiles, it provides a positive-semidefinite representation of subject-level morphometric similarity. The diagonal was retained, and matrices were symmetrized before trace normalization.

\subsection{Resting-state fMRI dataset: ABIDE}
\label{supp:abide}

Resting-state fMRI data were obtained from the ABIDE Preprocessed initiative \cite{dimartino2014abide,craddock2013neurobureau}. Subjects labelled as autism spectrum disorder or control were retained, yielding 871 subjects: 403 individuals with autism spectrum disorder and 468 controls. Regional haemodynamic time series were extracted using the AAL atlas \cite{tzourio2002aal}, resulting in 116 anatomical regions. For each subject, pairwise Pearson correlations between regional time series were computed to obtain a $116 \times 116$ functional-coupling matrix.

The ABIDE dataset is multi-site, and site-related variability is therefore an intrinsic characteristic of this empirical regime \cite{abraham2017multisite}. In the present study, ABIDE was used to test whether the prediction--explanation invariance problem generalizes to a large, heterogeneous functional-connectivity dataset.

\subsection{EEG dataset: OpenNeuro ds004504}
\label{supp:eeg}

EEG data were obtained from OpenNeuro ds004504 \cite{miltiadous2023eeg}. For the main analysis, subjects labelled as Alzheimer's disease and cognitively normal were retained, while the frontotemporal dementia group available in the original dataset was excluded. This yielded 65 subjects: 36 patients with Alzheimer's disease and 29 cognitively normal controls. EEG recordings were average-referenced, band-pass filtered between 1 and 45~Hz, and resampled to 250~Hz. For each subject, pairwise Pearson correlations between 19 scalp EEG channels were computed, yielding a $19 \times 19$ electrophysiological-coupling matrix \cite{bastos2016connectivity,vandiessen2015eeg}.

\subsection{Common matrix checks and standardization}
\label{supp:standardization}

After dataset-specific matrix generation, all matrices were processed using the same standardization pipeline. Matrices were symmetrized, their positive-semidefinite status was assessed by eigendecomposition, and trace normalization was applied to obtain unit-trace matrices. No thresholding or sparsification was applied before geometric, predictive, or perturbation-based analyses. Small negative eigenvalues attributable to numerical precision were tolerated when negligible relative to matrix scale.

\begin{table}[H]
\centering
\scriptsize
\setlength{\tabcolsep}{4pt}
\renewcommand{\arraystretch}{1.15}

\caption{Dataset-level summary}
\label{tab:s1}

\begin{tabularx}{\textwidth}{
@{}
P{1.6cm}
P{1.6cm}
P{1.3cm}
P{1.8cm}
c
P{1.6cm}
Y
P{1.8cm}
@{}}
\toprule

\textbf{Dataset} &
\textbf{Source} &
\textbf{Task} &
\textbf{Subjects} &
\textbf{Nodes} &
\textbf{Node definition} &
\textbf{Matrix construction} &
\textbf{Main caveat} \\

\midrule

\makecell[l]{Morph.\\MRI} &
ADNI &
AD vs CTRL &
\makecell[l]{96 AD\\119 CTRL} &
549 &
\makecell[l]{Spatial\\patches} &
\makecell[l]{Pearson similarity\\between z-normalized\\ patch intensity profiles} &
\makecell[l]{Patch-level\\anatomical\\interpretation} \\

fMRI &
\makecell[l]{ABIDE\\Preprocessed} &
ASD vs CTRL &
\makecell[l]{403 ASD\\468 CTRL} &
116 &
\makecell[l]{AAL\\regions} &
\makecell[l]{Pearson correlation\\between regional\\haemodynamic\\time series} &
\makecell[l]{Multi-site\\heterogeneity} \\

EEG &
\makecell[l]{OpenNeuro\\ds004504} &
AD vs CN &
\makecell[l]{36 AD\\29 CN} &
19 &
\makecell[l]{Scalp\\channels} &
\makecell[l]{Pearson correlation\\between channel\\time series} &
\makecell[l]{Low-dimensional\\channel\\representation} \\

\bottomrule
\end{tabularx}
\end{table}

\section{Geometric definitions and perturbation sensitivity}
\label{supp:geometry}

This note formalizes the geometric objects used in the main analysis. The purpose is to make explicit how empirical relation matrices are converted into trace-normalized positive-semidefinite representations, how Frobenius and trace geometries are computed, and how node perturbations generate geometry-dependent explanatory rankings.

\subsection{Empirical relation matrices}
\label{supp:relation-matrices}

For each subject $s$, the empirical data were represented by a symmetric relation matrix
\begin{equation}
A^{(s)} \in \mathbb{R}^{n \times n},
\end{equation}
where $n$ denotes the number of nodes in the corresponding representation. Nodes were morphometric patches for MRI, anatomical regions for fMRI, and scalp channels for EEG. Matrix entries encoded pairwise empirical relations between nodes: patch-profile similarity, haemodynamic coupling, or electrophysiological coupling.

The analyses require matrices that support a well-defined geometry. In the present datasets, the subject-level matrices were positive semidefinite within numerical tolerance. This allows each matrix to be interpreted as a Gram, kernel, or covariance-like object with a non-negative spectrum \cite{bhatia2007positive,scholkopf2002learning}.

\subsection{Spectral descriptors and edge-wise features}
\label{supp:representation-features}

Two feature-based representations were derived directly from each
trace-normalized matrix $\rho^{(s)}$: compact spectral descriptors and
edge-wise density features.

Let
\[
\lambda_1^{(s)} \geq \lambda_2^{(s)} \geq \cdots
\geq \lambda_n^{(s)} \geq 0
\]
denote the eigenvalues of $\rho^{(s)}$, arranged in descending order.
Because $\rho^{(s)}$ has unit trace,
\[
\sum_{j=1}^{n}\lambda_j^{(s)}=1.
\]

The compact spectral representation included the von Neumann entropy
\begin{equation}
S\!\left(\rho^{(s)}\right)
=
-\sum_{\lambda_j^{(s)}>10^{-12}}
\lambda_j^{(s)}\log \lambda_j^{(s)},
\end{equation}
the purity
\begin{equation}
\mathcal{P}\!\left(\rho^{(s)}\right)
=
\operatorname{Tr}\!\left[\left(\rho^{(s)}\right)^2\right]
=
\sum_{j=1}^{n}\left(\lambda_j^{(s)}\right)^2,
\end{equation}
and the effective rank
\begin{equation}
r_{\mathrm{eff}}\!\left(\rho^{(s)}\right)
=
\exp\!\left(S\!\left(\rho^{(s)}\right)\right).
\end{equation}

The spectral gap was defined as the difference between the two largest
eigenvalues,
\begin{equation}
\Delta\lambda^{(s)}
=
\lambda_1^{(s)}-\lambda_2^{(s)}.
\end{equation}
The feature vector additionally contained the first 20 eigenvalues and
the cumulative spectral masses
\begin{equation}
M_k^{(s)}
=
\sum_{j=1}^{k}\lambda_j^{(s)},
\qquad
k\in\{5,10,20\}.
\end{equation}

For representations containing fewer than 20 nodes, all available
eigenvalues were retained and cumulative spectral masses exceeding the
matrix dimension were omitted. Consequently, the morphometric MRI and
fMRI spectral representations contained 27 features: four global
spectral descriptors, the first 20 eigenvalues and three cumulative
spectral masses. The EEG representation contained 25 features: four
global descriptors, all 19 eigenvalues and the cumulative masses for
$k=5$ and $k=10$.

Edge-wise density features were obtained by vectorizing the strictly
upper-triangular, off-diagonal entries of each trace-normalized matrix.
This representation retained all pairwise node relations without
introducing an explicit subject-space metric. Its dimensionality was
$n(n-1)/2$, corresponding to 150426 features for morphometric MRI,
6670 features for fMRI and 171 features for EEG.

\subsection{Trace-normalized density representation}
\label{supp:density-representation}

For each subject $s$, the corresponding matrix $A^{(s)}$ was converted into a trace-normalized operator:
\begin{equation}
\rho^{(s)} = \frac{A^{(s)}}{\operatorname{Tr}\!\left[A^{(s)}\right]}.
\label{eq:supp-density}
\end{equation}
Because the empirical matrices had unit diagonal, $\operatorname{Tr}[A^{(s)}]$ was equal to the number of nodes $n$ before normalization. The trace-normalized representation therefore satisfies
\begin{equation}
\rho^{(s)} \succeq 0,
\qquad
\operatorname{Tr}\!\left[\rho^{(s)}\right]=1.
\end{equation}
This common representation makes it possible to compare otherwise heterogeneous empirical matrices within the same geometric framework \cite{nielsen2010quantum}.

\subsection{Frobenius geometry}
\label{supp:frobenius}

The Frobenius distance between two trace-normalized matrices $\rho^{(s)}$ and $\rho^{(t)}$ was defined as
\begin{equation}
d_{\mathrm{F}}(s,t)
= \left\|\rho^{(s)}-\rho^{(t)}\right\|_{\mathrm{F}}
= \sqrt{\operatorname{Tr}\!\left[\left(\rho^{(s)}-\rho^{(t)}\right)^2\right]}.
\label{eq:supp-frobenius}
\end{equation}
For real symmetric matrices, this distance corresponds to the Euclidean norm of the matrix difference. It therefore emphasizes distributed entry-wise deviations between subject-level relation matrices.

\subsection{Trace geometry}
\label{supp:trace}

The trace distance between two trace-normalized matrices $\rho^{(s)}$ and $\rho^{(t)}$ was defined as
\begin{equation}
d_{\mathrm{T}}(s,t)
= \frac{1}{2}\left\|\rho^{(s)}-\rho^{(t)}\right\|_{1}
= \frac{1}{2}\operatorname{Tr}\!\left|\rho^{(s)}-\rho^{(t)}\right|.
\label{eq:supp-trace}
\end{equation}
Because $\rho^{(s)}-\rho^{(t)}$ is Hermitian, its trace norm equals the sum of the absolute values of its eigenvalues. This geometry therefore compares matrices through operator-level directions along which they can be distinguished. In quantum information theory, the same expression is a standard distinguishability measure for density operators \cite{nielsen2010quantum}; here it is used operationally as a geometry on trace-normalized positive-semidefinite empirical matrices.

\subsection{Subject-space distance matrices and neighbourhoods}
\label{supp:subject-space}

For each geometry $\mathcal{G} \in \{\mathrm{F},\mathrm{T}\}$, pairwise distances were computed between all subjects, producing a subject-space distance matrix $D_\mathcal{G}$ with entries $d_\mathcal{G}(s,t)$. Subject-space concordance between Frobenius and trace geometries was assessed by Spearman correlation between the upper-triangular entries of the two distance matrices \cite{kriegeskorte2008rsa}. Local neighbourhood overlap was assessed by comparing the $k$ nearest neighbours of each subject under the two distance matrices using the Jaccard index.

\subsection{Node perturbations}
\label{supp:perturbations}

Let $P_i$ denote the leave-node-out perturbation associated with node $i$. Applying $P_i$ suppresses the contribution of all matrix entries incident to node $i$, namely those belonging to its corresponding row and column. The same node perturbation was applied under both geometries, with node $i$ corresponding to a morphometric patch, an fMRI region or an EEG channel depending on the empirical regime.

For a geometry $\mathcal{G} \in \{F,T\}$, let $d_\mathcal{G}(s,t)$ denote the original distance between held-out subject $s$ and training subject $t$, and let $d_\mathcal{G}^{(-i)}(s,t)$ denote the corresponding distance after suppressing node $i$. The pairwise geometric change can be written as

\begin{equation}
\Delta_{i,\mathcal{G}}(s,t) = d_\mathcal{G}^{(-i)}(s,t) - d_\mathcal{G}(s,t)
\end{equation}.

Its magnitude $|\Delta_{i,\mathcal{G}}(s,t)|$ describes how strongly node suppression modifies the pairwise geometry. This quantity is an intermediate geometric effect and is not the final node-importance score.

\subsection{Propagation through the kernel decision function}\label{supp:propagation}

For each outer cross-validation split, a support-vector machine was fitted using a precomputed radial-basis-function kernel derived from the selected geometry. The unperturbed kernel value between held-out subject $s$ and training subject $t$ was

\begin{equation}
    K_{\mathcal{G}}(s,t) = \exp{[-\gamma_{\mathcal{G}}d_{\mathcal{G}}(s,t)^2]}
\end{equation}

whereas node suppression produced

\begin{equation}
    K_{\mathcal{G}}^{(-i)}(s,t) = \exp{[-\gamma_{\mathcal{G}}d_{\mathcal{G}}^{(-i)}(s,t)^2]}
\end{equation}

The fitted classifier and its hyperparameters were kept fixed. Only the kernel row associated with the held-out subject was recomputed after node suppression.

Writing the fitted SVM decision function as

\begin{equation}
    f_{\mathcal{G}}(s) = \sum_{t\in SV}\beta_t K_{\mathcal{G}}(s,t) + b
\end{equation}

where $SV$ denotes the support-vector set, $b$ denotes the intercept (bias) of the fitted support-vector machine and $\beta_t$ incorporates the corresponding dual coefficient and class sign, the node-suppressed decision score was

\begin{equation}
    f_{\mathcal{G}}^{(-i)}(s) = \sum_{t\in SV}\beta_t K_{\mathcal{G}}^{(-i)}(s,t) + b
\end{equation}

Subject-specific node importance was defined as

\begin{equation}
    I_{i,\mathcal{G}}(s) = |f_{\mathcal{G}}(s) - f_{\mathcal{G}}^{(-i)}(s)|.
\end{equation}

Equivalently,

\begin{equation}
    I_{i,\mathcal{G}}(s) = \left|\sum_{t \in SV} \beta_t \left[K_{\mathcal{G}}(s,t)-K_{\mathcal{G}}^{(-i)}(s,t)\right]\right|.
\end{equation}

Thus, node importance is a support-vector-weighted aggregation of the kernel changes induced by suppressing an interpretable node of the original matrix representation.

\subsection{Frobenius node suppression}\label{supp:frob_suppr}
Let 

\begin{equation*}
   X_{st}
 = \rho^{(s)} - \rho^{(t)} 
\end{equation*}

The squared Frobenius distance is

\begin{equation}
    d_F(s,t)^2 = \sum_{j,k} (X_{st})_{jk}^2 
\end{equation}

The contribution involving node $i$ is

\begin{equation}
    c_{i,F}(s,t) = \sum_j (X_{st})_{ij}^2 + \sum_j (X_{st})_{ji}^2 - (X_{st})_{ii}^2
\end{equation}

The diagonal term is subtracted once because it appears in both the row and column sums. The node-suppressed Frobenius distance used in the implementation was

\begin{equation}
    d_F^{(-i)}(s,t) = \sqrt{\max\left[d_F(s,t)^2 - c_{i,F}(s,t),0\right]}.
\end{equation}

This is the exact Frobenius distance obtained after suppressing all entries incident to node $i$, without reconstructing the full reduced matrix.

\subsection{First-order trace-distance suppression}\label{supp:first_order}

Let 

\begin{equation*}
   X_{st}
 = \rho^{(s)} - \rho^{(t)} 
\end{equation*}

and

\begin{equation}
    d_T(s,t) = \frac{1}{2}||X_{st}||_1.
\end{equation}

If 

\begin{equation*}
    X_{st} = U\Lambda U^{T},
\end{equation*}

its matrix sign is

\begin{equation}
    sign(X_{st}) = U sign(\Lambda) U^{T}.
\end{equation}

Define 

\begin{equation}
    C_{st} = sign(X_{st}) \odot X_{st}
\end{equation}

where $\odot$ denotes the element-wise product. The first-order contribution associated with node $i$ was computed as

\begin{equation}
    c_{i,T}^{(1)(s,t)} = \frac{1}{2}\left[\sum_j (C_{st})_{ij} + \sum_j (C_{st})_{ji} - (C_{st})_{ii} \right].
\end{equation}

The node-suppressed trace distance was approximated by

\begin{equation}
    d_T^{(-i)}(s,t) \simeq max\left[ d_T(s,t)-c_{i,T}^{(1)(s,t)},0 \right].
\end{equation}

The approximated distance was then propagated through the radial-basis kernel and the fixed SVM decision function as in Eqs. S13–S17. In the morphometric MRI dataset, the first-order approximation was validated against exact leave-patch-out trace-distance recomputation for the highest-ranked trace-relevant patches. Exact and first-order mean absolute perturbation effects were strongly concordant at the patch level (Pearson $r=0.996$; Spearman $r=1.000$). Across subject pairs within the selected patches, Spearman correlations ranged from $0.987$ to $0.999$, with complete sign agreement. The approximation tended to overestimate effect magnitude, but preserved ranking and sign.

\subsection{Global and group-difference importance}
\label{supp:importance}

Global node importance was obtained by averaging the subject-specific decision-score changes across held-out subjects and cross-validation evaluations:

\begin{equation}
I_{i,\mathcal{G}}
= \frac{1}{M}
\sum_{s=1}^{M}I_{i,\mathcal{G}}(s).
\label{eq:supp-global-importance}
\end{equation}

Group-difference importance was the signed contrast

\begin{equation}
I_{i,\mathcal{G}}^{\mathrm{diff}}
= I_{i,\mathcal{G}}^{+}-I_{i,\mathcal{G}}^{-},
\label{eq:supp-diff-importance}
\end{equation}

where class-specific profiles were defined as

\begin{align}
I_{i,\mathcal{G}}^{+}
&= \frac{1}{|C_{+}|}
\sum_{s\in C_{+}}
I_{i,\mathcal{G}}(s),
\\[3pt]
I_{i,\mathcal{G}}^{-}
&= \frac{1}{|C_{-}|}
\sum_{s\in C_{-}}
I_{i,\mathcal{G}}(s).
\end{align}

In the three empirical analyses, the positive class was Alzheimer's disease for morphometric MRI, autism spectrum disorder for fMRI, and Alzheimer's disease for EEG.  Frobenius- and trace-based explanation-rank concordance was quantified using Spearman and Pearson correlations between the corresponding node-importance profiles. Top-$k$ overlap and consensus-ranking analyses were used as complementary descriptive summaries. The resulting quantities are geometry-conditioned explanations of classifier decisions: perturbations are defined on interpretable nodes of the original positive-semidefinite representation, while their relevance is quantified from the induced change in the decision function of a classifier operating directly on the corresponding geometry.

\section{Predictive performance across classifiers and representations}
\label{supp:ml}

To determine whether the observed differences between compact spectral descriptors and full-matrix representations depended on the choice of learning algorithm, we examined classifier-specific performance within the same repeated nested cross-validation framework used for the main analyses. Compact spectral descriptors were evaluated using L2-regularized logistic regression, linear support-vector machines, radial-basis-function support-vector machines and Random Forests. Edge-wise density features were evaluated using L2-regularized logistic regression and linear support-vector machines. Frobenius- and trace-based representations were evaluated using support-vector machines with precomputed radial-basis kernels derived from the corresponding pairwise distance matrices. All performance estimates were obtained from 100 outer held-out evaluations generated by $20$ repetitions of stratified five-fold nested cross-validation.

In morphometric MRI, compact spectral descriptors showed limited predictive performance across classifier families. The best spectral model was the Random Forest, with a mean ROC-AUC of $0.588 \pm 0.076$, followed by L2-regularized logistic regression with $0.567 \pm 0.076$. Linear and RBF-kernel support-vector machines yielded ROC-AUC values of $0.482 \pm 0.098$ and $0.429 \pm 0.106$, respectively. The weak performance of the spectral representation was therefore not restricted to a particular classifier. By contrast, edge-wise density features achieved closely matched ROC-AUC values with linear SVM and L2-regularized logistic regression, reaching $0.831 \pm 0.058$ and $0.829 \pm 0.058$, respectively. Frobenius- and trace-based kernel classifiers also achieved nearly identical discrimination, with ROC-AUC values of $0.825 \pm 0.069$ and $0.826 \pm 0.065$.

Despite their closely matched ROC-AUC values, the two geometric classifiers showed modest differences in class-specific operating characteristics. The Frobenius-based classifier achieved a sensitivity of $0.672 \pm 0.109$ and a specificity of $0.831 \pm 0.086$, whereas the trace-based classifier achieved a lower sensitivity of $0.632 \pm 0.106$ and a higher specificity of $0.853 \pm 0.078$. Thus, predictive discrimination was nearly invariant across the two geometries, although their operating points were not identical.

In fMRI, all classifiers applied to spectral descriptors performed close to chance. Mean ROC-AUC ranged from $0.472 \pm 0.031$ for Random Forests to $0.499 \pm 0.038$ for linear SVM, with L2-regularized logistic regression and RBF-kernel SVM yielding $0.499 \pm 0.038$ and $0.485 \pm 0.038$, respectively. In contrast, edge-wise density features produced almost identical performance with linear SVM and L2-regularized logistic regression, with ROC-AUC values of $0.711 \pm 0.042$ for both models. Frobenius- and trace-based classifiers achieved ROC-AUC values of $0.709 \pm 0.044$ and $0.718 \pm 0.039$, respectively.

The fMRI geometry-based classifiers also displayed class-specific imbalances that were not evident from ROC-AUC alone. The Frobenius classifier achieved a sensitivity of $0.570 \pm 0.060$ and a specificity of $0.732 \pm 0.055$, whereas the trace classifier achieved a sensitivity of $0.543 \pm 0.087$ and a specificity of $0.759 \pm 0.060$. Thus, trace geometry showed slightly higher overall discrimination but also a more pronounced bias towards specificity.

In EEG, spectral descriptors again showed limited predictive performance across classifiers. Mean ROC-AUC values were $0.517 \pm 0.148$ for L2-regularized logistic regression, $0.509 \pm 0.146$ for linear SVM, $0.500 \pm 0.149$ for Random Forests and $0.451 \pm 0.153$ for RBF-kernel SVM. Edge-wise density features retained substantially more predictive information, achieving ROC-AUC values of $0.703 \pm 0.125$ with linear SVM and $0.690 \pm 0.135$ with L2-regularized logistic regression. Frobenius- and trace-based classifiers yielded ROC-AUC values of $0.698 \pm 0.138$ and $0.688 \pm 0.136$, respectively.

In EEG, the geometry-based classifiers showed a different class-wise imbalance from that observed in morphometric MRI and fMRI. Frobenius geometry yielded a sensitivity of $0.811 \pm 0.203$ and a specificity of $0.360 \pm 0.289$, whereas trace geometry yielded a sensitivity of $0.836 \pm 0.185$ and a specificity of $0.322 \pm 0.309$. Their ROC-AUC values therefore remained comparable to the edge-wise benchmark, but their predictions were more strongly oriented towards the Alzheimer’s disease class.

Overall, the reduced performance of compact spectral descriptors was consistent across classifier families and therefore cannot be attributed solely to the use of an unsuitable learning algorithm. The close agreement between linear SVM and L2-regularized logistic regression for edge-wise density features further supports the robustness of the full-matrix benchmark. Frobenius- and trace-based classifiers retained broadly comparable discriminative performance across all datasets, while secondary metrics revealed modest and dataset-specific differences in sensitivity and specificity. These results support the main conclusion that predictive information was largely preserved across the two geometries, without implying complete invariance of their class-specific operating characteristics.

\scriptsize
\setlength{\tabcolsep}{2pt}
\renewcommand{\arraystretch}{1.12}

\begin{tabularx}{\linewidth}{
    >{\raggedright\arraybackslash}p{1.7cm}
    >{\raggedright\arraybackslash}p{2.2cm}
    >{\raggedright\arraybackslash}p{2.8cm}
    *{5}{>{\centering\arraybackslash}p{1.55cm}}
}

\caption{
\textbf{Classifier-specific predictive performance across representations and datasets.}
Predictive performance is reported as mean $\pm$ standard deviation across
100 outer held-out evaluations generated by 20 repetitions of stratified
five-fold nested cross-validation. Compact spectral descriptors were
evaluated using L2-regularized logistic regression, linear support-vector
machines, RBF-kernel support-vector machines, and random forests.
Edge-wise density features were evaluated using L2-regularized logistic
regression and linear support-vector machines. Frobenius- and trace-based
representations were evaluated using support-vector machines with
precomputed RBF kernels derived from the corresponding pairwise distance
matrices. Sensitivity refers to the positive diagnostic class---Alzheimer's
disease for morphometric MRI and EEG, and autism spectrum disorder for
fMRI---whereas specificity refers to the corresponding control class.
}
\label{tab:classifier_specific_performance}
\\

\toprule
\textbf{Dataset} &
\textbf{Representation} &
\textbf{Classifier} &
\textbf{ROC-AUC} &
\textbf{Balanced accuracy} &
\textbf{Accuracy} &
\textbf{Sensitivity} &
\textbf{Specificity}
\\
\midrule
\endfirsthead

\multicolumn{8}{c}{
\tablename\ \thetable{} -- continued
}
\\
\toprule
\textbf{Dataset} &
\textbf{Representation} &
\textbf{Classifier} &
\textbf{ROC-AUC} &
\textbf{Balanced accuracy} &
\textbf{Accuracy} &
\textbf{Sensitivity} &
\textbf{Specificity}
\\
\midrule
\endhead

\midrule
\multicolumn{8}{r}{Continued on next page}
\\
\endfoot

\bottomrule
\endlastfoot

Morphometric MRI &
Spectral descriptors &
L2 logistic regression &
$0.567 \pm 0.076$ &
$0.549 \pm 0.063$ &
$0.553 \pm 0.063$ &
$0.512 \pm 0.103$ &
$0.586 \pm 0.097$
\\

Morphometric MRI &
Spectral descriptors &
Linear SVM &
$0.482 \pm 0.098$ &
$0.532 \pm 0.053$ &
$0.553 \pm 0.055$ &
$0.338 \pm 0.187$ &
$0.726 \pm 0.179$
\\

Morphometric MRI &
Spectral descriptors &
RBF-kernel SVM &
$0.429 \pm 0.106$ &
$0.512 \pm 0.040$ &
$0.554 \pm 0.035$ &
$0.128 \pm 0.228$ &
$0.897 \pm 0.187$
\\

Morphometric MRI &
Spectral descriptors &
Random forest &
$0.588 \pm 0.076$ &
$0.561 \pm 0.068$ &
$0.567 \pm 0.068$ &
$0.500 \pm 0.118$ &
$0.622 \pm 0.111$
\\

Morphometric MRI &
Edge-wise density features &
L2 logistic regression &
$0.829 \pm 0.058$ &
$0.746 \pm 0.064$ &
$0.747 \pm 0.063$ &
$0.731 \pm 0.106$ &
$0.760 \pm 0.083$
\\

Morphometric MRI &
Edge-wise density features &
Linear SVM &
$0.831 \pm 0.058$ &
$0.746 \pm 0.062$ &
$0.748 \pm 0.060$ &
$0.725 \pm 0.104$ &
$0.766 \pm 0.078$
\\

Morphometric MRI &
Frobenius geometry &
Precomputed RBF-kernel SVM &
$0.825 \pm 0.069$ &
$0.751 \pm 0.067$ &
$0.760 \pm 0.066$ &
$0.672 \pm 0.109$ &
$0.831 \pm 0.086$
\\

Morphometric MRI &
Trace geometry &
Precomputed RBF-kernel SVM &
$0.826 \pm 0.065$ &
$0.743 \pm 0.065$ &
$0.755 \pm 0.064$ &
$0.632 \pm 0.106$ &
$0.853 \pm 0.078$
\\

\midrule

fMRI &
Spectral descriptors &
L2 logistic regression &
$0.499 \pm 0.038$ &
$0.501 \pm 0.032$ &
$0.509 \pm 0.034$ &
$0.388 \pm 0.082$ &
$0.614 \pm 0.091$
\\

fMRI &
Spectral descriptors &
Linear SVM &
$0.499 \pm 0.038$ &
$0.503 \pm 0.033$ &
$0.513 \pm 0.034$ &
$0.367 \pm 0.067$ &
$0.639 \pm 0.074$
\\

fMRI &
Spectral descriptors &
RBF-kernel SVM &
$0.485 \pm 0.038$ &
$0.498 \pm 0.019$ &
$0.520 \pm 0.026$ &
$0.201 \pm 0.194$ &
$0.794 \pm 0.199$
\\

fMRI &
Spectral descriptors &
Random forest &
$0.472 \pm 0.031$ &
$0.470 \pm 0.028$ &
$0.478 \pm 0.028$ &
$0.358 \pm 0.062$ &
$0.582 \pm 0.056$
\\

fMRI &
Edge-wise density features &
L2 logistic regression &
$0.711 \pm 0.042$ &
$0.656 \pm 0.039$ &
$0.657 \pm 0.038$ &
$0.650 \pm 0.062$ &
$0.662 \pm 0.056$
\\

fMRI &
Edge-wise density features &
Linear SVM &
$0.711 \pm 0.042$ &
$0.657 \pm 0.039$ &
$0.658 \pm 0.039$ &
$0.649 \pm 0.061$ &
$0.666 \pm 0.057$
\\

fMRI &
Frobenius geometry &
Precomputed RBF-kernel SVM &
$0.709 \pm 0.044$ &
$0.651 \pm 0.037$ &
$0.657 \pm 0.037$ &
$0.570 \pm 0.060$ &
$0.732 \pm 0.055$
\\

fMRI &
Trace geometry &
Precomputed RBF-kernel SVM &
$0.718 \pm 0.039$ &
$0.651 \pm 0.039$ &
$0.659 \pm 0.037$ &
$0.543 \pm 0.087$ &
$0.759 \pm 0.060$
\\

\midrule

EEG &
Spectral descriptors &
L2 logistic regression &
$0.517 \pm 0.148$ &
$0.513 \pm 0.113$ &
$0.518 \pm 0.115$ &
$0.570 \pm 0.187$ &
$0.456 \pm 0.192$
\\

EEG &
Spectral descriptors &
Linear SVM &
$0.509 \pm 0.146$ &
$0.504 \pm 0.113$ &
$0.511 \pm 0.114$ &
$0.569 \pm 0.187$ &
$0.439 \pm 0.190$
\\

EEG &
Spectral descriptors &
RBF-kernel SVM &
$0.451 \pm 0.153$ &
$0.487 \pm 0.100$ &
$0.510 \pm 0.106$ &
$0.719 \pm 0.265$ &
$0.256 \pm 0.257$
\\

EEG &
Spectral descriptors &
Random forest &
$0.500 \pm 0.149$ &
$0.506 \pm 0.134$ &
$0.515 \pm 0.133$ &
$0.579 \pm 0.200$ &
$0.434 \pm 0.225$
\\

EEG &
Edge-wise density features &
L2 logistic regression &
$0.690 \pm 0.135$ &
$0.649 \pm 0.117$ &
$0.648 \pm 0.116$ &
$0.639 \pm 0.180$ &
$0.659 \pm 0.189$
\\

EEG &
Edge-wise density features &
Linear SVM &
$0.703 \pm 0.125$ &
$0.664 \pm 0.110$ &
$0.665 \pm 0.107$ &
$0.669 \pm 0.170$ &
$0.659 \pm 0.191$
\\

EEG &
Frobenius geometry &
Precomputed RBF-kernel SVM &
$0.698 \pm 0.138$ &
$0.585 \pm 0.105$ &
$0.608 \pm 0.098$ &
$0.811 \pm 0.203$ &
$0.360 \pm 0.289$
\\

EEG &
Trace geometry &
Precomputed RBF-kernel SVM &
$0.688 \pm 0.136$ &
$0.579 \pm 0.101$ &
$0.608 \pm 0.088$ &
$0.836 \pm 0.185$ &
$0.322 \pm 0.309$
\\

\end{tabularx}

\normalsize

\section{Classifier definitions and implementation details}
\label{supp:classifiers}

All predictive analyses were implemented in Python using NumPy
($= 1.26$), pandas ($= 2.1$), SciPy ($= 1.11$),
scikit-learn ($= 1.4$), joblib ($= 1.3$), and Matplotlib
($= 3.8$). NumPy and SciPy were used for numerical and matrix
operations, pandas for data organization, scikit-learn for model
implementation, preprocessing, cross-validation and performance
evaluation, joblib for parallel computation, and Matplotlib for
visualization.

Hyperparameters were selected within the inner loop of the nested
cross-validation procedure. Model performance was subsequently evaluated
on the corresponding held-out outer folds. For models requiring feature
standardization, scaling and classification were combined within a
scikit-learn \texttt{Pipeline}. The mean and standard deviation used for
standardization were therefore estimated exclusively from the training
data of each cross-validation fold, preventing information leakage from
the validation or test samples.

\subsection*{L2-regularized logistic regression}
Logistic regression is a linear probabilistic classifier that models the
log-odds of class membership as a weighted combination of the input
features. The decision boundary is therefore linear in the original
feature space. An L2 penalty was applied to the model coefficients to
reduce coefficient magnitude and improve generalization, particularly
when the number of features was large relative to the number of subjects.

Before model fitting, features were standardized using
\texttt{StandardScaler}. Optimization was performed using the
\texttt{liblinear} solver, with a maximum of 5000 iterations. The inverse
regularization parameter $C$ was selected from
\[
C \in
\left\{
10^{-4},10^{-3},10^{-2},10^{-1},10^{0},10^{1},10^{2},10^{3}
\right\}.
\]
Smaller values of $C$ correspond to stronger regularization, whereas
larger values allow the classifier to fit the training data more closely.

\subsection*{Linear support-vector machine}
The linear support-vector machine identifies a separating hyperplane that
maximizes the margin between the two diagnostic classes. Unlike logistic
regression, its optimization objective is based on the hinge loss and
focuses primarily on samples located close to the decision boundary. The
model produces continuous decision scores describing the signed distance
of each sample from the separating hyperplane.

Features were standardized before classification. The maximum number of
optimization iterations was set to 10000, and the
\texttt{dual} parameter was set to \texttt{auto}, allowing
scikit-learn to select the appropriate primal or dual optimization
formulation according to the relationship between the number of samples
and features. The penalty parameter was optimized over

\[
C \in
\left\{
10^{-4},10^{-3},10^{-2},10^{-1},10^{0},10^{1},10^{2},10^{3}
\right\}.
\]
As in logistic regression, smaller values of $C$ impose stronger
regularization, while larger values penalize classification errors more
strongly.

\subsection*{Radial-basis-function support-vector machines}

Radial-basis-function support-vector machines were used in two distinct
ways, depending on the representation being analysed. For the compact
spectral descriptors, the RBF kernel was computed directly between
explicit feature vectors. For the Frobenius- and trace-geometry
representations, the RBF kernel was instead precomputed from the
corresponding subject-to-subject distances. Although both approaches use
the same Gaussian functional form, they differ in what is compared and in
how the kernel-width parameter is selected.

\paragraph{RBF kernel applied to explicit spectral features.}

For compact spectral descriptors, each subject was represented by an
explicit feature vector $\mathbf{x}$. The similarity between two subjects
with feature vectors $\mathbf{x}$ and $\mathbf{x}'$ was defined as

\[
K(\mathbf{x},\mathbf{x}')
=
\exp\left(
-\gamma \lVert \mathbf{x}-\mathbf{x}' \rVert_2^2
\right).
\]

The kernel therefore measured the Euclidean similarity between subjects
in the standardized spectral-feature space. This nonlinear mapping
allowed the classifier to construct nonlinear decision boundaries without
explicitly computing the corresponding transformed features.

Spectral features were standardized before model fitting. Scaling and
classification were included in the same scikit-learn pipeline, so that
the standardization parameters were estimated exclusively from the
training data within each cross-validation fold.

Two hyperparameters were optimized. The penalty parameter was selected
from

\[
C \in
\left\{
10^{-3},10^{-2},10^{-1},10^{0},10^{1},10^{2},10^{3}
\right\},
\]

while the kernel-width parameter was selected from

\[
\gamma \in \{\texttt{scale},\texttt{auto}\}.
\]

Under the scikit-learn implementation, \texttt{scale} sets

\[
\gamma
=
\frac{1}{p\,\mathrm{Var}(X)},
\]

whereas \texttt{auto} sets

\[
\gamma
=
\frac{1}{p},
\]

where $p$ is the number of input features. Larger values of $\gamma$
produce more localized decision boundaries, whereas smaller values
generate smoother boundaries influenced by a broader range of training
samples.

\paragraph{Precomputed RBF kernel applied to Frobenius and trace distances.}

The Frobenius- and trace-geometry classifiers did not operate on explicit
feature vectors. Instead, each geometry directly defined the distance
between subjects. Let

\[
d_g(s,t),
\qquad
g \in \{\mathrm{F},\mathrm{T}\},
\]

denote the Frobenius or trace distance between the trace-normalized
matrices of subjects $s$ and $t$. For each geometry, the corresponding
distance was transformed into a Gaussian similarity:

\[
K_g(s,t)
=
\exp\left(
-\gamma d_g(s,t)^2
\right).
\]

The resulting matrix $K_g$ was supplied to the support-vector machine
using the scikit-learn \texttt{kernel=precomputed} interface. Thus, the
distance matrix itself was not treated as a feature matrix and was not
used directly as the kernel. Rather, the SVM received the similarity
matrix obtained by applying the RBF transformation to the
geometry-specific distances.

Within each outer cross-validation split, the kernel used to fit the
classifier was constructed only between training subjects:

\[
K_{g,\mathrm{train}}(s,t)
=
\exp\left(
-\gamma d_g(s,t)^2
\right),
\qquad
s,t\in\mathcal{T},
\]

where $\mathcal{T}$ denotes the outer training set. For held-out
evaluation, the test kernel contained similarities between each test
subject and the outer-training subjects:

\[
K_{g,\mathrm{test}}(s,t)
=
\exp\left(
-\gamma d_g(s,t)^2
\right),
\qquad
s\in\mathcal{V},\quad t\in\mathcal{T},
\]

where $\mathcal{V}$ denotes the corresponding outer test set. The fitting
kernel therefore had dimensions
$n_{\mathrm{train}}\times n_{\mathrm{train}}$, whereas the held-out
kernel had dimensions
$n_{\mathrm{test}}\times n_{\mathrm{train}}$.

For these geometry-based models, the penalty parameter was selected from

\[
C \in
\left\{
10^{-3},10^{-2},10^{-1},10^{0},10^{1},10^{2},10^{3}
\right\}.
\]

The candidate kernel widths were determined separately for each geometry
and within each outer training set. Let

\[
m_g^2
=
\operatorname{median}
\left\{
d_g(s,t)^2:
s<t,\;
s,t\in\mathcal{T},\;
d_g(s,t)>0
\right\}
\]

denote the median of the non-zero squared distances among the outer
training subjects. A reference value was defined as

\[
\gamma_{0,g}
=
\frac{1}{m_g^2},
\]

and the candidate kernel widths were

\[
\gamma
\in
\gamma_{0,g}
\left\{
0.01,\,
0.03,\,
0.1,\,
0.3,\,
1,\,
3,\,
10
\right\}.
\]

Both $C$ and $\gamma$ were selected by maximizing ROC-AUC within the
inner cross-validation loop. Hyperparameter selection was performed
independently for the Frobenius and trace geometries because the two
metrics generally produced distance matrices with different numerical
scales and distributions. Feature standardization was not applied to
these models because their inputs were precomputed kernel matrices rather
than explicit feature vectors.

\subsection*{Random forest}
The random forest is a nonlinear ensemble classifier that combines the
predictions of multiple decision trees. Each tree recursively partitions
the feature space using feature-specific decision rules, while the final
classification is obtained by aggregating the predictions across trees.
The combination of multiple randomized trees reduces the variance
associated with an individual decision tree and allows the model to
capture nonlinear relationships and interactions between features.

Each forest contained 300 trees. The number of candidate features
considered at each split was selected from
\[
\texttt{max\_features}
\in
\left\{
\sqrt{p},\,0.3p,\,0.6p
\right\},
\]
where $p$ denotes the total number of input features. The minimum number
of samples required in each terminal leaf was selected from
\[
\texttt{min\_samples\_leaf} \in \{1,2,5\}.
\]
Increasing this parameter restricts the formation of leaves containing
very few samples and therefore provides additional regularization.

Class weights were set to \texttt{balanced}, so that the contribution of
each diagnostic class was weighted inversely according to its frequency
in the training data. The random seed was fixed according to the common
analysis configuration to ensure reproducibility, and all available
processing cores were used during model fitting through
\texttt{n\_jobs=-1}. Feature standardization was not applied because
decision-tree splitting rules are invariant to monotonic rescaling of the
individual features.

\addcontentsline{toc}{section}{Supplementary references}
\label{supp:references}

\bibliographystyle{unsrt}
\bibliography{apssamp}